%% file: paper.tex
\documentclass[journal]{IEEEtran}
\IEEEoverridecommandlockouts%
\usepackage{cite}
\usepackage{amsmath,amssymb,amsfonts}
\usepackage{algorithmic}
\usepackage{graphicx}
\usepackage{textcomp}
\usepackage{xcolor}
\usepackage[nolist]{acronym}
\usepackage{import}
\usepackage{graphicx} 
\usepackage{multicol} 
\usepackage{multirow}
\usepackage{array}
\usepackage{hyperref}
\usepackage{pgfplots}
\usepackage{pgfplotstable}
\usepackage{tikz}
\usepackage{subfigure}
\usepackage{subfigure}
\usepackage{colortbl}
\usepackage{soul}
\usepackage{float}
\floatstyle{plaintop}
\restylefloat{table}
\usepackage{comment}
\usepackage[bottom]{footmisc}
\usepackage{hyperref}
\usepackage{xspace}

\usepackage{url}

\tikzstyle{startstop} = [rectangle,  minimum width=3cm, minimum height=1cm,text centered, draw=black]
\tikzstyle{stomp}=[rectangle,minimum width=3cm, minimum height=3cm, text centered, draw=black]
\tikzstyle{stop} = [rectangle, minimum width=2cm, minimum height=1cm, draw=black]
\tikzstyle{big_rectangle} = [rectangle, minimum width=20.5cm, minimum height=7cm, draw=black, thick, dashed]
\tikzstyle{detail_rect} = [rectangle, minimum width=11cm, minimum height=5cm, draw=black]
\tikzstyle{non_dot_lined_rectangle} = [rectangle, minimum width=20.5cm, minimum height=10.5cm, draw=black, thick]
\tikzstyle{io} = [trapezium, trapezium left angle=70, trapezium right angle=110, minimum width=3cm, minimum height=1cm, text centered, draw=black, fill=blue!30]
\tikzstyle{process} = [rectangle, minimum width=3cm, minimum height=1cm, text centered, draw=black, fill=orange!30]
\tikzstyle{decision} = [diamond, minimum width=3cm, minimum height=1cm, text centered, draw=black, fill=green!30]
\tikzstyle{arrow} = [thick,->,>=stealth]
\tikzstyle{dual_arrow} = [thick, <->]
\tikzstyle{no_arrow}=
[thick]

\usepackage[justification=centering]{caption}
\usepackage[ruled,vlined]{algorithm2e}
\usepackage{amsmath,amssymb,amsfonts}
\usepackage{algorithmic}
\usepackage{algorithm2e}
\usepackage{graphicx}
\usepackage{xspace}
\usepackage{textcomp}
\usepackage[nolist]{acronym}
\usepackage{color, colortbl}
\usepackage{adjustbox}
\renewcommand{\thetable}{\arabic{table}}
\usepackage[colorinlistoftodos]{todonotes}
\usepackage{pgfplots}
\usepackage{pgfplotstable}
\usepackage{tikz}
\usepackage{comment}
\usepackage[nolist]{acronym}
\usepackage{blindtext}
\usepackage{changepage}
\usepackage{url}
\usepackage{soul}
\usepackage[english]{babel}
\usepackage[utf8x]{inputenc}
\usepackage[T1]{fontenc}
\usepackage[numbers]{natbib}
\usepackage{pdfpages}
\usepackage{listings}
\usepackage{hyperref}
\usepackage{lscape}
\usepackage{colortbl}
\usepackage{multirow}
\usepackage{multicol, lipsum}
\usepackage{diagbox}
\usepackage{adjustbox}
\usepackage{booktabs}
\usepackage[colorinlistoftodos]{todonotes}
\usepackage{enumitem}
\usepackage{lscape}
\usepackage{tikz}
\usepackage{xspace}

\usepackage[justification=centering]{caption}

\renewcommand{\thetable}{\arabic{table}}

\def\BibTeX{{\rm{}B\kern-.05em{\sc{}i\kern-.025em b}\kern-.08em
    T\kern-.1667em\lower.7ex\hbox{E}\kern-.125emX}}

\input{Sections/acro}
\pgfplotsset{compat=1.17}
\begin{document}

\title{Challenges in the Design and Implementation of IoT Testbeds in Smart-Cities: A Systematic Review}

\author{Vijay~Kumar,
        Sam~Gunner$^*$,
        Theodoros~Spyridopoulos$^*$,
        Antonis~Vafeas$^*$,
        James~Pope,
        Poonam~Yadav,
        George~Oikonomou, and~Theo~Tryfonas
\thanks{V. Kumar, S.Gunner and T.Tryfonas are with the Department of Civil Engineering, University of Bristol, UK, e-mail: first.lastname@bristol.ac.uk.}
\thanks{J. Pope and G. Oikonomou are with the Department of Engineering Mathematics and Electrical and Electronics Engineering, University of Bristol, UK respectively, e-mail: first.lastname@bristol.ac.uk.}
\thanks{T. Spyridopoulos is with the School of Computer Science and Informatics, Cardiff University, UK, e-mail: spyridopoulost@cardiff.ac.uk.}
\thanks{P. Yadav is with Computer Science Department, University of York, UK, e-mail: poonam.yadav@york.ac.uk.}
\thanks{A. Vafeas is with Computer Science Department, University of Bristol, UK, e-mail: vafeas.2011@my.bristol.ac.uk.}
\thanks{$^*$ S. Gunner, T Spyridopoulos and A. Vafeas contributed equally to this work}}

\markboth{Working draft - under review - 2023}%
{Kumar \MakeLowercase{\textit{et al.}}: Bare Demo of IEEEtran.cls for IEEE Journals}

\maketitle

\begin{abstract}
    \input{Sections/0_Abstract.tex}
\end{abstract}


\section{Introduction}
\label{sec:introduction}
\input{Sections/1_Introduction.tex}

\section{Background}
\label{sec:background}
\input{Sections/2_Background.tex}

\section{Literature Review}
\label{sec:literature_review}
\input{Sections/3_Literature_Review.tex}

\section{Methodology}
\label{sec:Methodology}
\input{Sections/4_Methodology.tex}

\section{Challenges}
\label{sec:challenges}
\input{Sections/5_Challenges.tex}

\section{Conclusion}
\label{sec:conclusion}
\input{Sections/6_conclusion.tex}

\section*{Acknowledgment}
This work has been supported in part by EPSRC through grant no EP/P016782/1 (UKCRIC Urban Observatories) and an Industrial CASE award sponsored by BT.

\bibliographystyle{./bibliography/IEEEtran}
\bibliography{./bibliography/paper_bib.bib}

\end{document}

%% file: Sections/acro.tex
\begin{acronym}
\acro{6LoWPAN}{IPv6 over Low Power Wireless Personal Area Networks}
\acro{SCK}{Smart Citizen Kit}
\acro{eCO2}{equivalent CO2}
\acro{UO}{Urban Observatory}
\acro{IoT}{Internet of Things}
\acro{WSN}{Wireless Sensor Networks}
\acro{SHM}{Structural Health Monitoring}
\acro{BSI}{British Standards Institute}
\acro{ICT}{Information and Communications Technology}
\acro{SolarPV}{Solar Photovoltaic Systems}
\acro{KPI}{Key Performance Indicators}
\acro{MCU}{Microcontrollers}
\acro{APNR}{Automatic Plate Number Recognition}
\acro{CCTV}{Closed-circuit television}
\acro{CO}{Carbon Monoxide}
\acro{CO2}{Carbon Dioxide}
\acro{NO}{Nitric Oxide}
\acro{NO2}{Nitrogen Dioxide}
\acro{CH4}{Methane}
\acro{H2S}{Hydrogen sulfide}
\acro{NH3}{Ammonia}
\acro{RSSI}{Received Signal Strength Indicator}
\acro{ECG}{Electrocardiogram}
\acro{LIDAR}{Light Detection and Ranging}
\acro{NOAA}{National Oceanic and Atmospheric Administration}
\acro{LTE}{Long Term Evolution}
\acro{GPRS}{General Packet Radio Service}
\acro{CDMA}{Code-division multiple access}
\acro{GSM}{Global System for Mobile Communication}
\acro{NB-IoT}{Narrowband IoT}
\acro{M2M}{Machine to Machine}
\acro{WiMAX}{Worldwide Interoperability for Microwave Access}
\acro{UWB}{Ultra Wideband}
\acro{BUO}{Bristol Urban Observatory}
\acro{LAN}{Local Area Network}
\acro{RF}{Radio-frequency}
\acro{RFID}{Radio-frequency identification}
\acro{NFC}{Near-field communication}
\acro{BCG}{Ballistocardiographic}
\acro{PIR}{Passive Infrared Sensor}
\acro{IAQ}{Indoor Air Quality}
\acro{VOC}{Volatile Organic Compounds}
\acro{EV}{Electric Vehicle}
\acro{TfL}{Transport for London}
\acro{TERS}{Temporary earth restraining structure}
\acro{SLA}{Service level agreement}
\acro{LoRaWAN}{Long Range Wide Area Network}
\acro{WiMAX}{Worldwide Interoperability for Microwave Access}
\acro{AoT}{Array of Things}
\acro{SBC}{Single Board Computer}
\acro{SPHERE}{Sensor Platform for HEalthcare in a Residential Environment}
\acro{SHG}{SPHERE Home Gateway}
\acro{SDH}{SPHERE Data Hub}
\acro{NUC}{Next Unit of Computing}
\acro{LCD}{Liquid Crystal Display}
\acro{SONET}{Synchronous optical networking}
\acro{DSL}{Digital Subscriber Line}
\acro{LED}{Light Emitting Diode}
\acro{SoC}{System on Chip}
\acro{TBI}{Traumatic Brain Injury}
\acro{AURN}{Automatic Urban and Rural Network}
\acro{VOCS}{Volatile Organic Compounds}
\acro{QoL}{Quality of Life}
\acro{AQI}{Air Quality Index}
\acro{UKCRIC}{UK Collaboratorium for Research On Infrastructure and Cities}
\acro{LWM2M}{Lightweight Machine to Machine}
\acro{OTA}{Over-the-Air}
\acro{JTAG}{Joint Test Action Group}
\acro{KWMC}{Knowle West Media Centre}
\acro{BCC}{Bristol City Council}
\acro{CAN}{Controller Area Network}
\acro{SBC}{single-board computer}
\acro{GPS}{Global Positioning System}
\acro{UART}{Universal Asynchronous Receiver/Transmitter}
\acro{PII}{Personally Identifiable Information}
\acro{STEM}{Science, Technology, Engineering and Mathematics}
\acro{VM}{virtual machine}
\acro{OS}{Operating System}
\acro{SD}{Secure Digital}
\acro{VPN}{Virtual Private Network}
\acro{DMZ}{Demilitarized Zone}
\acro{SDP}{Software Defined Perimeter}
\acro{SIEM}{Security Incident Event Monitoring}
\acro{COTS}{commercial off-the-shelf}
\acro{GUI}{Graphical User Interface}
\acro{LPWAN}{low-power wide-area network}
\acro{SSH}{Secure Shell}
\acro{TPM}{Trusted Platform Module}
\acro{NVD}{National Vulnerability Database}
\acro{IPR}{Ingress Protection Ratings}
\acro{PM}{Particulate Matter}
\acro{UMA}{Urban monitoring architecture}
\acro{OEM}{original equipment manufacturer}
\acro{NTP}{Network Time Protocol}
\acro{HDD}{Hard disk drive}
\acro{GDPR}{General Data Protection Regulation}
\acro{MITM}{man-in-the-middle}
\acro{VUI}{voice-user interface}
\acro{eMMC}{Embedded MultiMediaCard}
\acro{TRL}{Technology Readiness Level}
\acro{V2I}{Vehicle-to-Infrastructure}
\acro{AR}{Augmented Reality}
\acro{BLE}{Bluetooth Low Energy}
\acro{USB}{Universal Serial Bus}
\acro{PDR}{packet delivery rate}
\acro{API}{Application Programming Interface}
\acro{SSD}{solid-state drive}
\acro{KISS}{Keep it simple, stupid}
\acro{PoE}{Power-over-Ethernet}
\acro{HVAC}{Heating, Ventilation, and Air Conditioning}
\acro{SLIP}{Serial to IP}
\acro{WAN}{Wide Area Network}
\acro{URL}{Uniform Resource Locator}
\acro{CPU}{Central Processing Unit}
\acro{I2C}{Inter-Integrated Circuit}
\acro{SPI}{ Serial Peripheral Interface}
\acro{RAN}{Radio Access Network}
\acro{CE}{Conformite Europeenne}
\acro{UDP}{user datagram protocol}
\acro{TCP}{transmission control protocol}
\acro{HTTP}{Hypertext Transfer Protocol}
\acro{TEE}{trusted execution environment}
\acro{PCB}{Printed Circuit Board}
\acro{JSON}{JavaScript Object Notation}
\acro{MQTT}{Message Queuing Telemetry Transport}
\acro{LDAP}{Lightweight Directory Access Protocol}
\acro{LUKS}{Linux Unified Key Setup}
\acro{PDR}{packet delivery ratio}
\acro{QR}{Quick Response code}
\acro{PCA}{principal component analysis}
\acro{GPU}{graphics processing unit}
\end{acronym}

%% file: Sections/0_Abstract.tex
Advancements in wireless communication and the increased accessibility to low-cost sensing and data processing IoT technologies have increased the research and development of urban monitoring systems. Most smart city research projects rely on deploying proprietary IoT testbeds for indoor and outdoor data collection. Such testbeds typically rely on a three-tier architecture composed of the Endpoint, the Edge, and the Cloud. Managing the system's operation whilst considering the security and privacy challenges that emerge, such as data privacy controls, network security, and security updates on the devices, is challenging. This work presents a systematic study of the challenges of developing, deploying and managing urban monitoring testbeds, as experienced in a series of urban monitoring research projects, followed by an analysis of the relevant literature. By identifying the challenges in the various projects and organising them under the V-model development lifecycle levels, we provide a reference guide for future projects. Understanding the challenges early on will facilitate current and future smart-cities IoT research projects to reduce implementation time and deliver secure and resilient testbeds.

%% file: Sections/1_Introduction.tex
\IEEEPARstart{U}{rbanization} has raised the demand for natural resources in cities, while increased pollution has also increased environmental impacts. As cities grow, the logistics to ensure the provision of essential services becomes more challenging for city councils~\cite{Revision2018, khan2013cloud}. 
To improve city management and allow the development of relevant services, councils monitor city parameters such as air quality, road traffic, pedestrian movement, electricity usage, etc. Emerging technologies such as \ac{IoT} provide the ability to understand the physical environment with more granular data and allow citizens and city councils to make better decisions. The opportunities offered by \ac{IoT} implementations are numerous. For example, monitoring \ac{VOCS} and \ac{eCO2} in households can help residents with long-term lung conditions (such as asthma) identify poor air quality and act accordingly.

City councils, in association with researchers and universities, participate in multiple research projects such as SPHERE~\cite{woznowski2017sphere, elsts2020tsch}, REPLICATE~\cite{replicate_website} and Twinergy~\cite{twinergy_website} that aim to improve energy use, mobility, human well-being and productivity, reduce energy footprint, and increase resilience~\cite{Kumar2023} and sustainability of the city~\cite{Catlett2017}. 

The \ac{AoT} team~\cite{Catlett2017} has conducted various workshops with multi-disciplinary academics and citizen communities to understand how IoT technology comprising sensors, cameras, and computation capabilities can help modern cities. They concluded that scientific instruments (endpoint/edge IoT devices) deployed in an urban environment to provide spatial and temporal sensor data for analysis could benefit residents and city councils. Their emerging IoT platform ultimately forms an urban-scale instrument for research and development~\cite{Catlett2017}, simultaneously testing new sensors, communication, and computing devices. 

Edge devices deployed in public spaces can also be used to test and support new technologies such as \ac{V2I} communication in Co-operative is Intelligent Transportation Systems and \ac{AR} to display city information. 

However, the development and management of urban monitoring systems pose many challenges. The collection of citizen data can lead to privacy violations if they are not properly managed~\cite{Kurkovsky2017}. The complexity of such systems and the integration of heterogeneous and in many cases, proprietary technologies further increase the data management problem and can also result in security issues that may ultimately disrupt services~\cite{6392021, 6513686, 8897627}. 

This paper aims to systematically identify the challenges in developing urban monitoring IoT testbeds based on the authors' experience in relevant \ac{UO}, smart city projects, and the analysis of the relevant literature. These projects include: Harbourside water quality monitoring~\cite{coraggiosmart, CHEN2018307}, Clifton Suspension Bridge structural health monitoring~\cite{gunner2017rapid}, e-bike monitoring~\cite{internet:e_bike}, damp residential detection~\cite{nepomuceno2019residential} and \ac{SCK} deployment in the Cotham Hill Pedestrianisation Programme, as well as others~\cite{sphere_website,replicate_website,twinergy_website}. \autoref{table:authors_involved} lists the different research projects in which the authors were involved and provided their experiences. \autoref{table:authors_referred} lists similar smart city projects that the authors referred to understand the challenges faced in the projects mentioned in the research publications. We hope this work will benefit the design and implementation of future smart cities research projects and IoT testbeds, reducing the implementation time.

\input{Tables/author_project_sync.tex}

\input{Tables/projects_referred_sync.tex}

The rest of the paper is structured as follows:
\autoref{sec:background} provides the background knowledge of smart cities research projects, testbed and monitoring architecture. \autoref{sec:literature_review} provides a brief about literature review. \autoref{sec:Methodology} provides the methodology followed to understand the challenges. \autoref{sec:challenges} provides the challenges faced by the research projects mapped to the V-model. \autoref{sec:conclusion} concludes the work.

%% file: Tables/author_project_sync.tex
\begin{table*}[]
\makebox[1 \textwidth][c]{       
\resizebox{1 \textwidth}{!}{  

\begin{tabular}{llllllll}
\hline
\textbf{Project} & \textbf{Size (Where)}    & \multicolumn{1}{l}{\textbf{Data collected}} & \multicolumn{5}{l|}{\textbf{Architecture}}               \\ \hline
SPHERE                            & (100 Homes), (1 EN; multiple EP)/home & Environmental                                                 & EP (802.15.4)      & $\rightarrow$ & EN(4G)          & $\rightarrow$ & CS \\
UMBRELLA                          & 200 EN (streetlamps) with on-board EP & Environmental, Camera                                         & EP (I2C/SPI)       & $\rightarrow$ & EN (Fibre/WiFi) & $\rightarrow$ & CS \\
Cotham Hill Pedestrianization     & 10 EP in (8 homes)                    & Noise and air pollution                                       & EP (WiFi)          & $\rightarrow$ &                 &               & CS \\
Residential Dampness              & (1 home), (1 EN with on-board EP)     & Temperature, Humidity                                         & EP (Analog)        & $\rightarrow$ & EN              &               &    \\
Clifton Suspension Bridge         & 1 EN, 2 EP                            & Structural health monitoring data                             & EP (802.15.4)      & $\rightarrow$ & EN (4G)         & $\rightarrow$ & CS \\
Water quality monitoring          & 3 sites (1 device with 7 sensors)     & Water quality                                                 & EP (Serial to WiFi) & $\rightarrow$ &                 & $\rightarrow$ & CS \\
SYNERGIA                          & 3 ENs, 15 EP (office)                 & Environmental                                                 & EP (802.15.4/LoRa) & $\rightarrow$ & EN (LAN)        & $\rightarrow$ & CS \\
REPLICATE (Energy)                & Smart appliances (151 Homes);         & Energy consumption                                            & EP (LAN)           & $\rightarrow$ & EN (LAN)        & $\rightarrow$ & CS \\
REPLICATE (eBike)                 & EN (12 e-bikes)                       & Battery level, motor power                                    & EP (CAN)           & $\rightarrow$ & EN (LoRa/WiFi)  & $\rightarrow$ & CS \\
Bristol AoT                       & 3 EN with on-board EP                 & Environmental, Camera                                         & EP (I2C/SPI)       & $\rightarrow$ & EN (4G)         & $\rightarrow$ & CS \\
Twinergy                          & 12 home                               & Energy consumption data                                       & EP (WiFi)          & $\rightarrow$ &                 & $\rightarrow$ & CS \\
EurValve                          & (40 homes), (4 EN; 1 EP)/home         & RSSI and accelerometer data                                   & EP (Bluetooth)     & $\rightarrow$ & EN(4G/WiFi)     & $\rightarrow$ & CS \\ \hline
\end{tabular}
} 
}
\caption{Research projects in which the authors participated. CS: Cloud Server, EN: Edge Node (Gateway), EP: Endpoints (IoT Node). CS may contain all or a subset of open-source components (Kafka, K3S, MQTT, InfluxDB, Grafana). EN may consist of SBC (RPi or Intel NUC)}
\label{table:authors_involved}
\end{table*}

%% file: Tables/projects_referred_sync.tex
\begin{table*}[]
\makebox[1 \textwidth][c]{       
\resizebox{1 \textwidth}{!}{

\begin{tabular}{llllllll}
\hline
\textbf{Project}                                        & \textbf{Size (Where)}        & \textbf{Data collected}    & \multicolumn{5}{l}{\textbf{Architecture}}                                            \\ \hline
AoT~\cite{Catlett2017}                             & 130 EN (streetlamps) with on-board EP         & Environmental, Camera                       & EP (I2C/SPI)               & $\rightarrow$ & EN (4G)                             & $\rightarrow$ & CS \\
e-Agriculture~\cite{Oyedele2021}                   & EN (Lab deployment)                           & Light, temperature, soil pH and humidity    & EP (Analogue)              & $\rightarrow$ & EN                                  &               &    \\
Living Labs~\cite{Jackson2017}                     & 150 EN, 800 EP (120 location)                 & Air quality, microclimating, bat monitoring & EP (RPL)                   & $\rightarrow$ & EN (2G)                             & $\rightarrow$ & CS \\
Connected Vehicle Testbed~\cite{Chowdhury2018}     & 3 Fixed EN (FEN), 2 Mobile EN (MEN)           & Vehicle position data                       & MEN (wireless)             & $\rightarrow$ & FEN (wired)                         & $\rightarrow$ & CS \\
Wireless Environmental Sensors~\cite{Folea2020}    & 1 EN, 7 EP (Lab deployment)                   & Environmental                               & EP (Bluetooth)             & $\rightarrow$ & EN (LAN)                            & $\rightarrow$ & CS \\
Solar-powered WSN~\cite{Dehwah2015}                & 82 EP (real-world deployment)                 & Temperature, RSSI, battery level            & EP (WSN) w/ sink           & $\rightarrow$ &                                     &               &    \\
Community Elderly Care~\cite{Valera2018}           & EN, EP (70 elderly homes)                     & Motion, door contact                        & EP (Z-wave)                & $\rightarrow$ & EN (cellular)                       & $\rightarrow$ & CS \\
IEEE802.15.4 Connectivity Traces~\cite{Traces2015} & 350 EP (Office environment)                   & RSSI, PDR                                   & EP (802.15.4) w/ sink      & $\rightarrow$ &                                     &               &    \\
LOFAR-Agro~\cite{Langendoen2006}                   & 109 EP, 3 EN, (real-world deployment)         & Temperature, humidity                       & EP (WSN) w/ sink           & $\rightarrow$ & EN (WiFi)                           & $\rightarrow$ & CS \\
3E Houses~\cite{Porto2013}                         & (100 homes)(6 EP/ 1 EN)/home                  & Energy consumption data                     & EP (Zigbee)                & $\rightarrow$ & EN (WiFi)                           & $\rightarrow$ & CS \\
New York Noise sensor network~\cite{Mydlarz2019}   & 55 Nodes (1 EP and 1 EN)/node                 & Noise data                                  & EP (USB)                   & $\rightarrow$ & EN (WiFi)                           & $\rightarrow$ & CS \\
Padova Smart City~\cite{6918931}                   & 1 EN, 8 EP                                    & Temperature, humidity, benzene              & EP (802.15.4)              & $\rightarrow$ & EN (WiFi)                           & $\rightarrow$ & CS \\
Flash Flood Monitoring~\cite{Basnyat2020}          & 3 iter. of IoT device deployed; EN, EP        & Water levels                                & EP (USB)                   & $\rightarrow$ & EN (cellular)                       & $\rightarrow$ & CS \\
Smart Santander~\cite{Sotres2017}                  & 50+ EN, 700+ EP                               & Environmental                               & EP (802.15.4)              & $\rightarrow$ & EN (Wired/Wireless)                 & $\rightarrow$ & CS \\
City of Things~\cite{Latre2016}                    & 32 locations (1 node/location;multi-radio)    & Air quality, traffic monitoring, parking    & EP (WSN)                   & $\rightarrow$ & EN (multi-radio)                    & $\rightarrow$ & CS \\
SADMote~\cite{Elsts}                               & 5 EN, 12 EP                                   & Environmental                               & EP (WSN)                   & $\rightarrow$ & EN (WiFi)                           & $\rightarrow$ & CS \\
SensorScope~\cite{Hnat2011}                        & $\approx$6EN, each serving $\approx$100 EP    & Environmental                               & EP (WSN)                   & $\rightarrow$ & EN (GPRS)                           & $\rightarrow$ & CS \\
EpiFi~\cite{Lundrigan2019}                         & $\approx$ 18 locations (2 EP, 1EN)/location   & Environmental                               & EP (WSN/WiFi)              & $\rightarrow$ & EN (WiFi)                           & $\rightarrow$ & CS \\
Parking System~\cite{Huang2017}                    & 2 EP, 3 EN                                    & Parking, Light sensor                       & EP (Lora)                  & $\rightarrow$ & EN (Lora receiver)                  &               &    \\
Residential Sensing~\cite{Hnat2011}                & $\approx$ 20 homes $\approx$ 1200 EP          & Temperature, light, door                    & EP (Z-wave)                & $\rightarrow$ & EN (WiFi)                           & $\rightarrow$ & CS \\
Water consumption~\cite{Yang2015}                  & 30 homes, 1EN, $\approx$ 4 EP                 & Water consumption                           & EP (433MHz)                & $\rightarrow$ & EN (WiFi)                           & $\rightarrow$ & CS \\ \hline
\end{tabular}
} 
}
\caption{Research projects referred by the authors. Based on the details provided in the papers. CS: Cloud Server, EN: Edge Node, EP: Endpoints.}
\label{table:authors_referred}
\end{table*}

%% file: Sections/2_Background.tex
\subsection{Smart Cities Research Projects}

Multiple organisations collaborate with the city council to make a city smart and work on research projects to improve citizen lives and city council services. Smart city research projects can target different areas, for example, collecting environmental data to monitor air, noise, water pollution, residential dampness, energy monitoring, or structural health monitoring of buildings and bridges. \autoref{table:authors_involved} and \autoref{table:authors_referred} provide a list of innovative city projects, the data they collect, their architecture, and the size of the deployment. Multiple smart city research projects deploy testbeds to collect urban or citizen health data for different analyses. Major implementations occur in public places or citizens' homes. In public places, there have been multiple projects such as Smart Santander~\cite{smartsantander_website}, UMBRELLA~\cite{umbrella_website}, and \ac{AoT}~\cite{Catlett2017}, whereas projects such as SPHERE~\cite{woznowski2017sphere, elsts2020tsch} and REPLICATE~\cite{replicate_website} have deployed devices in citizen homes. Smart Santander deployed multiple IEEE 802.15.4 devices, \ac{GPRS} modules, and joint \ac{RFID} tag/\ac{QR} code labels deployed at both static locations (streetlights, facades, bus stops) and mobile vehicles (buses, taxis) for different smart city use cases. Similarly, UMBRELLA deployed multiple edge nodes mounted on lamp posts containing wireless radio nodes and sensors, providing a real-world platform to test wireless algorithms and smart city sensing (temperature, air quality, and noise). The AoT project deployed edge nodes in Chicago to collect real-time data on the city's environment, infrastructure, and activity for research and public use. SPHERE deployed a multi-modal platform of non-medical home sensors to serve as a prototype for future residential healthcare systems. REPLICATE deployed edge devices to deploy energy efficiency, mobility, and \ac{ICT} solutions in city districts. Twinergy has installed house batteries and smart plugs in people's houses to improve their self-consumption of locally generated renewable energy and monitor their uptake of energy demand side management.

\subsection{A Brief About Testbeds}

Testbeds play an essential role in experimental research by allowing researchers to perform experiments, deploy multiple devices, set up realistic environments, and collect sensor data and insights~\cite{Nussbaum2017, Yadav2017, Feraudo2020}. The testbeds are made up of endpoints (sensors that sense the physical parameter), edge gateways (collect and process data from endpoints) and cloud infrastructure (collect and process data from endpoints/edge). Managing such an infrastructure is challenging~\cite{Langendoen2006}. The challenges include the security and management of multiple devices, data security and privacy, user privacy controls, visualisation, multitenancy of applications, hardware malfunctions, programming bugs, software incompatibilities, network resilience, and plain misunderstanding of concepts~\cite{Langendoen2006, bellavista2019survey}. Furthermore, each research project implements the testbed differently based on the project team's requirements, usability, budget, time, and technical skillset.

A testbed should enable researchers to \textbf{i.} deploy and connect multiple devices at the edge and endpoint tier safely and securely, \textbf{ii.} deploy applications on the cloud and edge devices collecting and processing data from the endpoints (sensors) and sending it securely to the edge/cloud,  \textbf{iii.} manage the devices for accounting and administrator purposes \textbf{iv.} provide data visualisation and insights to end users~\cite{Lin} and \textbf{v.} be adaptable to fulfill other requirements. 
The authors categorised the testbed into three different categories  \textbf{i.} Distributed large-scale cloud resources testbed providing researchers the access to the bare metal and control over computing, storage, and networking resource, e.g., Chameleon~\cite{chameleon_website}, GENI~\cite{geni_website}, GRID5000~\cite{grid5000_website}, FED4FIRE/FED4FIRE+~\cite{fed4fire_website}, FIT-Cloud~\cite{fitcloud_website}, Emulab~\cite{emulab_website}, PlanetLab~\cite{planetlab_website}, PRAGMA~\cite{pragma_website}, DETER~\cite{deter_website}, NOR-NET Core~\cite{nornet_website}, SAVI~\cite{savi_website}  \textbf{ii.} distributed large-scale endpoint \ac{WSN} testbed that provide access to the WSN nodes to conduct network experiments, i.e., FIT IoT-Lab~\cite{fitiot_website}, SmartSantander~\cite{smartsantander_website}, City of Things~\cite{citylab_website}, UMBRELLA~\cite{umbrella_website}  \textbf{iii.} data-collecting research testbed that collects data from citizen house or public spaces, i.e., SPHERE~\cite{woznowski2017sphere, elsts2020tsch}, REPLICATE~\cite{replicate_website}, Twinergy~\cite{twinergy_website}, 3E Houses~\cite{3ehouses_website}, SONYC~\cite{Mydlarz2019}, \ac{AoT}~\cite{Catlett2017}, Scallop4SC~\cite{matsumotoscallop4sc}, Padova~\cite{6918931}.

\subsection{A General Three Tier Architecture}

\begin{figure}
    \centering
    \includegraphics[width=\columnwidth]{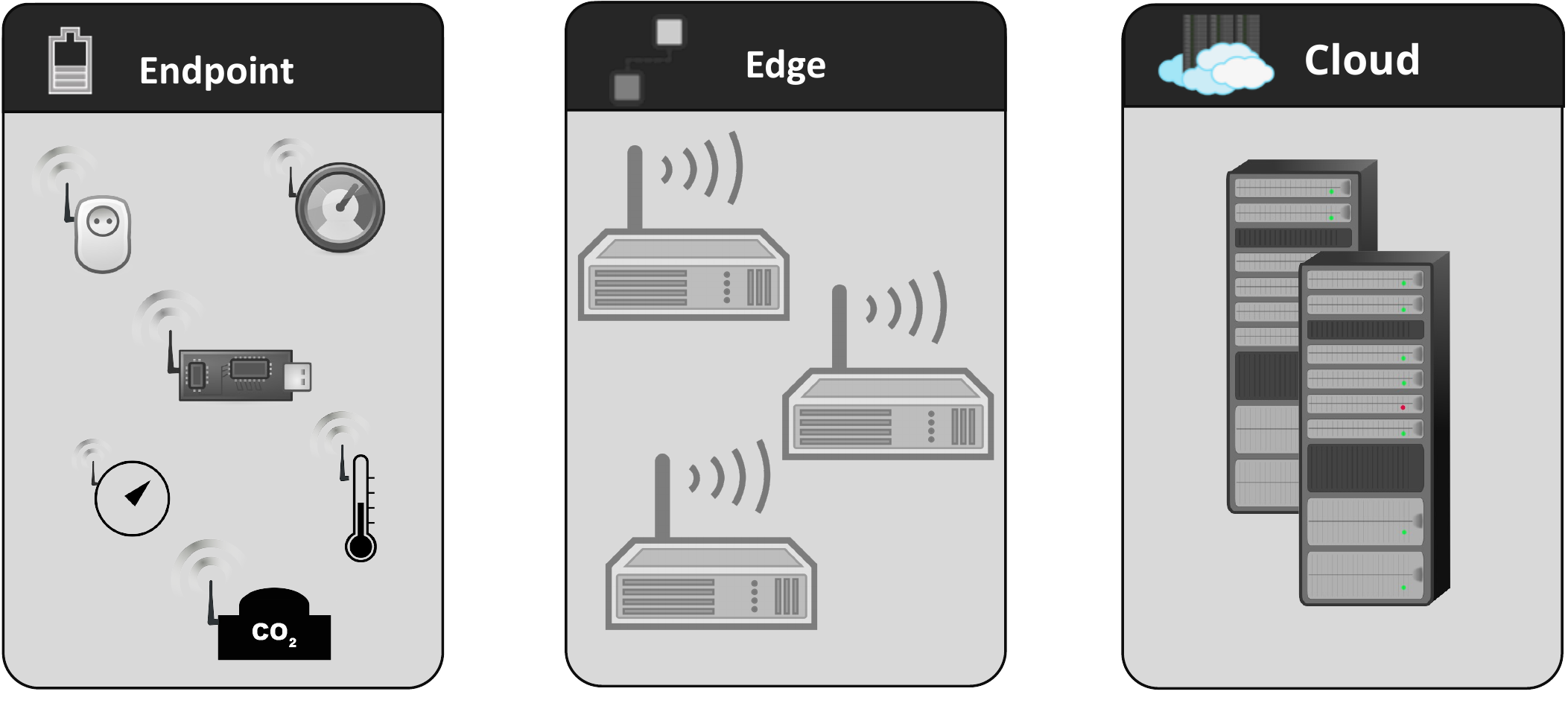}
    \caption[Background: Typical three-tier architecture for a smart city]{Typical three-tier architecture for a smart city}
    \label{fig:cloud_edge_endpoint}
\end{figure}

Testbeds can have different architectures based on the project requirements, such as endpoint-cloud, endpoint-edge, and endpoint-edge-cloud. In endpoint-cloud, devices at the endpoint tier communicate directly with the cloud tier; in endpoint-edge, the endpoint sends the data to the edge, and the cloud tier does not exist. In endpoint-edge-cloud, endpoints connect directly to edge devices, and devices at the edge tier connect to the cloud tier. Endpoint-edge-cloud is a standard architecture used by different projects such as SPHERE~\cite{woznowski2017sphere, elsts2020tsch}, REPLICATE~\cite{replicate_website}, Clifton Suspension Bridge~\cite{gunner2017rapid}, AoT~\cite{Catlett2017} and others~\cite{Basnyat2020, lee2010u} and also mentioned in relevant literature~\cite{9139976, bellavista2019survey}.

\autoref{fig:cloud_edge_endpoint} presents a typical architecture of a data collection testbed consisting of cloud, edge, and endpoint tiers. We provide a brief introduction about each tier below:

\subsubsection*{\textbf{Endpoint Tier}}
The endpoint tier consists of resource-constrained, battery-powered embedded devices with low-power wireless communications capability. The devices are generally inconspicuous and have a small nominal form factor for deployment in space-constrained environments~\cite{Bevers2019}. They can sense different environmental parameters such as barometric pressure, temperature, humidity, light, motion (with an accelerometer, gyroscope, or compass) and presence (using an infrared sensor to detect the human body's heat). In addition, a reed relay or switches can sense the opening/closing of a window/door. Endpoints generate monitoring data and send it to a collection point at the edge/cloud tier for processing and analysis. The endpoints can be connected to the edge/cloud tier by different technologies such as \textbf{i.} an IEEE 802.15.4 network (in a mesh or star topology) created and controlled by an edge tier device, \textbf{ii.} \ac{LPWAN} network technologies (Sigfox, LoRaWAN, NB-IoT, Wi-Fi, \ac{BLE}),  \textbf{iii.} directly connected to the edge device using a \ac{USB}, \ac{I2C}, \ac{SPI}, \ac{UART}.

\subsubsection*{\textbf{Edge Tier}}
The edge level can consist of a \ac{SBC} (Raspberry Pi (RPi), Jetson Nano (JN), Grapeboard, Intel NUC) installed in a citizen's home or public spaces (street lamps, bus stops, city council vehicles) or private buildings~\cite{Gvk2019}. The edge tier collects the data sent by the endpoints and either process it or sends it in a raw format to the cloud tier~\cite{Sanchez2014} for further analysis. Processing data at the edge reduces payload size and communication bandwidth, shortens latency, and simplifies data formatting and aggregation for the cloud~\cite{Beavers}. The edge device can also run different applications, such as urban environment monitoring and counting people/vehicles, and is often designed to be application-agnostic. It provides end users with a sensing/processing element at the network edge that can service novel applications. Edge devices are typically connected to the cloud tier using higher bandwidth and more reliable communications technologies, such as 4G/5G, Wi-Fi, and fibre. According to the project requirements, the edge device can contain multiple radios onboard, such as IEEE 802.1ac on 2.4/5 GHz, DASH7 on 433/868 Mhz, \ac{BLE}, IEEE 802.15.4, IEEE 802.15.4g, and LoRa~\cite{Latre2016}. Edge devices (based on their location) can also be used in infrastructure mode (endpoints connecting to edge tier) or ad hoc peer-to-peer (edge devices connecting to each other using radios).

\subsubsection*{\textbf{Cloud Tier}}
The cloud tier consists of multiple servers, hosting all the applications and services required to manage the devices at the edge, endpoint tier and the applications necessary to achieve the project objectives~\cite{Sanchez2014}. The servers will run multiple components on \ac{VM} or containers. The cloud tier can be hosted privately (OpenStack, VMWare) or on commercial cloud services (Azure, AWS). It contains the application logic and services required to operate and manage the testbed platform. The cloud tier should provide different services to the edge tier, such as credential management, data storage, provisioning of devices, networking, time synchronisation, secure remote software updating, configuring, and maintaining access to the edge and endpoint devices. It should also provide a secure communication channel to devices and services in the edge and endpoint tiers.

With the advancements in core networks, part of the functionality is distributed from the cloud tier to multiple geographical locations towards the edge network. In this case, the main point of distinction becomes \ac{RAN} and how the endpoint tier is connected to the cloud tier. Depending on the wireless and wired transmission network, some core network features, computation, and offloading can occur on the edge tier. Therefore, it is vital to address the challenges of edge-to-cloud connectivity and the architectural decisions that each testbed has chosen.

%% file: Sections/3_Literature_Review.tex
Santana et al. ~\cite{10.1145/3124391} surveyed multiple smart cities projects in the area of cyber-physical systems, IoT, Big Data and cloud computing. They provided challenges and open research problems in developing next-generation, robust software platforms for smart cities. They included privacy (data owners, data usage), data management (storing, processing a large amount of data and trusting it), heterogeneity (different devices, data sources), energy management (energy consumption failure), communication (network reliability), scalability (increase in the number of users, devices, data), security (safe from cyber-terrorism and cyber-vandalism), lack of testbeds, city models (effective and efficient city model), platform maintenance (mange devices).

There have been multiple lessons learned papers regarding deployment of battery-powered devices in the endpoint tier communicating over IEEE 802.15.4~\cite{Fafoutis2017, Elsts, Langendoen2006, Barrenetxea2008, Traces2015}, devices deployed in public spaces~\cite{Catlett2017, Cheng2015}, citizens houses~\cite{Valera2018, Fafoutis2017, Schleich2012, Elsts2017, beaudin2004lessons}, testbeds~\cite{Webb2004, Keahey2020}, and others~\cite{KumarR2019, Staudemeyer2019, Ekedahl2018, Palattella2016, In1998, Dehwah2015, Folea2020, Chowdhury2018, Jackson2017, Lago2021}. However, the lessons learnt are from a project with specific requirements such as "challenges in flash flood monitoring" or does not include human engagement (deployment in citizen houses or public spaces) or has very small-scale deployment. They do not cover every stage or all the challenges faced during a smart-city research project. Our work focuses on a three-tier architecture (cloud, edge and endpoints) usually used in smart-cities research projects. It covers the end-to-end perspective and the challenges faced during the research project, from the requirement analysis, system design, implementation, and integration testing to the final deployment stage, keeping security and scalability in mind. It is more expansive and covers a larger scope, providing learnings from deploying multiple smart-city research projects.

%% file: Sections/4_Methodology.tex
The methodology authors used to understand the challenges faced in the smart-cities research project is interview-based and literature review-based. To collect the necessary data for our research, we conducted semi-structured interviews with the system architects and the deployment team of European research projects on IoT platforms and testbeds for urban monitoring~\cite{nepomuceno2019residential, gunner2017rapid, coraggiosmart, sphere_website, replicate_website, twinergy_website}. \autoref{table:authors_involved} lists the different research projects authors were involved in and provides their experiences. The interviews focused on the challenges the participants faced during the development, deployment and management of the IoT platforms and testbeds. The author asked simple open-ended questions with a free-flowing approach by asking a set of questions to the interviewee, and the conversation was continued based on the answers. The questions asked were about the challenges faced, such as ``\textit{What are the challenges faced during the projects}'', ``\textit{How did we provision the devices}'', ``\textit{What was the architecture of the project}'', ``\textit{How did the devices communicate with each other}'', ``\textit{How did we manage the storage, credentials}'', ``\textit{Any challenges faced in the implementation, deployment}'', ``\textit{What could have made your (system architect) life better}'', ``\textit{any unexpected challenges}''. The authors captured additional challenges based on their reflections on their experiences as members of smart-cities projects.

Moreover, we thoroughly reviewed the relevant literature on infrastructure deployment. \autoref{table:authors_referred} provides a summary of different projects that authors referred to understand the challenges faced in projects deploying IoT infrastructure.

To facilitate the exploitation of our work by future projects, we categorized the identified challenges based on the stage of the project lifecycle they appear. Almost all engineering projects follow a similar development lifecycle, from ``requirement analysis'' and ``system design'' to ``integration and testing'' and final project delivery. In our work, we identify the challenges in the various project and organize them under the V-model's level to formalize the development process and provide a reference guide for future projects.

\subsection{The V-model}


The V-model is one of several project life cycle models developed over time. Project life-cycle models try to visualise and map the different stages of a technology development project. They are an essential tool for the engineer and provide a standard conceptual framework of reference~\cite{Forsberg2005-na}.

\begin{figure}[t!]
	\centering
	\includegraphics[scale=0.3]{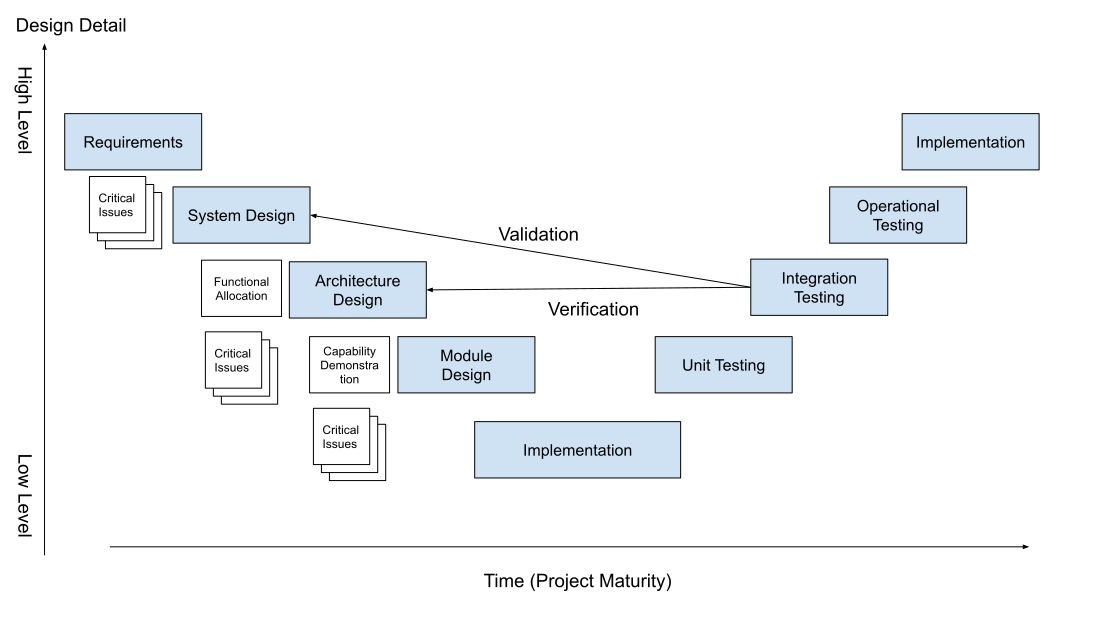}
	\caption{A graphical representation of the V-Model (modified from~\cite{Forsberg1991-cv}) }
	\label{fig:VeeModel}
\end{figure}

The V model~\cite{Rook1986-qk} is based heavily on `the waterfall model'~\cite{Royce1987-cv} that preceded it but increased it by projecting the project cycle into a three-dimensional space. \autoref{fig:VeeModel} shows the first two dimensions of this space, x and y, representing `time' (or project maturity) and `Design Detail', respectively. The `Design Detail' axis has high-level design at the top and low-level (or detailed) design decisions at the bottom. The central elements of the model (referred to as the core of the Vee) are shown in blue. The specific phraseology used in these elements varies depending on the particular application. However, the general theme is always the same: as the project works down the left arm of the V, high-level system designs are converted into more detailed system designs. 

One failure of the waterfall model was its embargo on any detailed design work before official approval of high-level design decisions~\cite{Forsberg1991-cv}. The V-Model removes this restriction, allowing the detailed technical enquiry to inform higher-level decisions. This ‘off-core’ work can be seen in the white boxes below the core of V. The work in the off-core varies depending on the design stage, but it aims to derisk the decisions currently being made. Important off-core work ~\cite{Forsberg1991-cv} is the identification of `critical issues'. Capability demonstrations are also important, demonstrating that technology can perform the desired functionality before it is written into a specification.

The final dimension of the V-Model, the z-axis pushing into the page, represents the different system design elements at that level of system decomposition. For example, architecture will consist of many modules, and each must be designed, so the V-model fans out to represent this, one branch for each module. Below the V, the z-axis represents the different and competing design options that must be evaluated before a selection can be made.
The workflow moves down the left-hand side of V until the bottom is reached, which means that design decisions are completed and can now be implemented. This means that each piece of hardware can be built and each software package written. 

Integrating these different components is necessary to form the final functioning system. It is performed by moving back on the right-hand side of the V. Each module is tested against the design from which it was created and then integrated with other modules to deliver more sophisticated functionality. That functionality is then tested against the higher-level design. Not only is ‘verification’ carried out (confirming that the module has been built according to its design specification), but ‘validation’ is also performed to ensure that the design captures the system's requirements.

%% file: Sections/5_Challenges.tex
\begin{figure*}
    \centering
    \includegraphics[width=\textwidth]{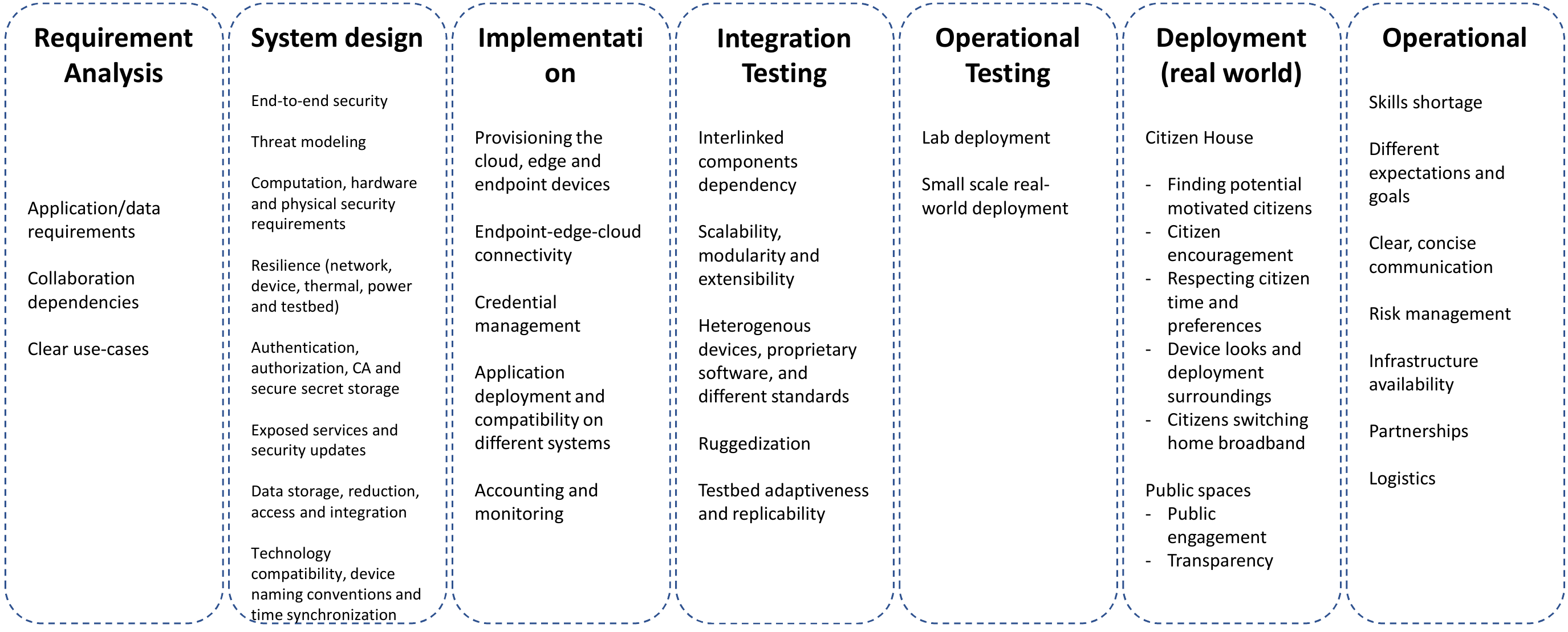}
    \caption[Challenges: Summary of challenges in smart-cities research projects]{Summary of challenges in smart-cities research projects}
    \label{fig:challenges_kumar}
\end{figure*}

This section discusses the challenges captured. We use the V-model to classify challenges and map various phases of the research project. \autoref{fig:challenges_kumar} summarises the challenges faced during different stages of the research project assigned to the V-model phases. Challenges can be categorised into multiple phases of a smart city research project, from understanding project requirements (requirement analysis) to designing how to fulfil those requirements (system design) and setting up defined infrastructure (implementation) to ensure that different infrastructure components work together (integration) and tested in the laboratory and initial small-scale deployment (operational testing) followed by deployment in the real world and operational challenges.

\subsection{Requirements Analysis}
\label{subsec:challenges:requirement}
The requirement analysis stage helps to understand the application and data requirements, collaboration dependency, and project use cases.

\subsubsection*{\textbf{Application/Data requirements}}
Data is at the heart of urban monitoring research projects. Depending on the need, it can be collected from multiple sensors deployed in citizens' houses, streetlamps, or bus stops. The nature of data required to meet project objectives and expected results affects, in general, all aspects of the project, from the technology to be used to the security implications of the privacy achieved~\cite{Chatterjee2018}. For example, in the SPHERE project, researchers created bespoke wearable devices with multiple components, many of which (e.g. second acceleration sensor, gyroscope, non-volatile flash memory, LED, button) were never used in real deployment~\cite{Fafoutis2017}. During the REPLICATE and Twinergy project, it was found that it is essential to engage with the stakeholders of the project (e.g., the city council and citizens) at the beginning of the project, clarify their expectations, understand their needs, and translate them into requirements for data collection, processing, storage, sharing, and visualisation~\cite{Gvk2019}.

Once the type of data is clarified, it is essential to consider the relevant \ac{GDPR}~\cite{EuropeanParliamentandtheCounciloftheEuropeanUnion2016} implications. In the UK, the Data Protection Act 2018 implements the European \ac{GDPR}. The Act introduces the terms ``data controller'' and ``data processor'' and clarifies the responsibilities around personal data collection, processing, and storage. These considerations will influence the system's design (e.g.\ employ mechanisms to ensure secure data collection, data anonymisation, or data destruction) and final deployment (e.g.\ deployment only after citizens' consent) in the subsequent development process steps. For example, the SPHERE project stored raw sensor data related to health in an external \ac{LUKS} encrypted~\ac{SSD}~\cite{Elsts2018}.

Great care must also be taken to ensure that the collected non-personal data cannot be used to infer information about individuals. For example, environmental/energy data can reveal citizens' behaviour and habits when not handled appropriately. Depending on the entities involved in the project, different actors may be interested in ensuring compliance with \ac{GDPR}. Universities undergo an ethical approval process that involves a rigorous analysis of relevant implications and solutions. City councils may require a privacy impact assessment that describes the data the project aims to collect, potential privacy issues, and the related impact.

In addition to legal implications around data collection, special care must be taken to clarify, understand, and comply with contractual agreements (e.g. data-sharing agreements) among the project's partners. The partnership agreement should detail the data each partner aims to collect, share, or process and the purpose of this activity, including potentially generated intellectual property and monetisation. This information should also be considered when considering the project's \ac{GDPR} implications.

Another data-related requirement that must be addressed in the early stages of a project is the need to integrate the data collected by the platform into other existing city data platforms (such as Bristol Open Data~\cite{bristol_open_data_website} and London Datastore~\cite{london_open_data_website}). Capturing integration requirements with external systems early on ensures the use of appropriate technologies and the timely delivery of the project.

Stakeholders must agree on the data requirements to ensure that the system's development follows user needs.

\subsubsection*{\textbf{Collaboration dependencies}}
Urban monitoring research projects often involve multiple partners (e.g. universities, city councils, industries) and require collaboration between different departments between partners (e.g. IT support, estate team). For example, project servers are usually behind the university or company firewall. The opening of ports on the firewall can take a considerable amount of time, ranging from weeks to months. The process may require multiple approvals from different entities and involves cyber security risk assessments to understand the various threats to the system and identify possible mitigation techniques. In projects with multiple collaborators, it is essential to consider these interactions and dependencies and address them during the requirements analysis period of the project.

\subsubsection*{\textbf{Clear use cases}}

Once the requirement collection has been completed, the project team must develop use cases that address the requirements~\cite{Gvk2019}. Below, we provide a few examples of use cases in urban monitoring projects:

\begin{itemize}
    \item{\textbf{Use case for deployment of sensors in Citizen Home (Indoor)}: Sensors provide details about indoor pollution and help citizens take action, such as opening windows for cross-ventilation.}
    \item{\textbf{Use case for deployment of sensors in a commercial building}: Assuming that a corporate building consists of multiple floors/rooms, the building management team can consist of a \ac{HVAC} team, estates teams, admin team, fire safety, and different companies occupying the offices/floor. Data can be sent to different teams depending on their requirements. For example, temperature data to the HVAC team to ensure the optimal temperature in rooms/offices; battery data to the estate's teams to ensure that the sensor batteries are replaced on time; air quality data and occupancy data to respective companies on respective floors.}
    \item{\textbf{Use case for deployment of sensors in public spaces}: When sensors are installed inside citizen homes, outdoor data (vehicle traffic, pedestrian traffic, light levels, weather, atmospheric conditions) can be compared with indoor data and provide context~\cite{Committee2018}. The sensor data can validate and train the various micro-climate weather models. Citizens can also use noise and air pollution data to decide on the suitability of buying a house in the neighbourhood}

\end{itemize}

\subsection{System Design}

%

The V-model system design phase provides a system overview, details of the different hardware, software, network protocols, applications, and logical components in the three-tier architecture mentioned in \autoref{sec:background} and the interfaces between them. It allows system architects to define testbed requirements from the perspectives of resources, security, resilience, data, and technology. System design decisions must be based on project requirements, and the requirements can always be referred back to understand and justify the design as specified in the V-model. For simplicity, the architecture and module design are merged into the system design.

\subsubsection*{\textbf{End-to-End Security}}
Securing a testbed from end-to-end (endpoint, edge, cloud tier) is challenging. It includes the security of all devices at each tier and the communication between them, including physical and data security. Endpoint-Edge-Cloud or End-to-End testbeds should be secure by design and provide fundamental security blocks such as confidentiality, integrity, availability, and non-repudiation~\cite{Staudemeyer2019, 7496795}. Confidentiality requires data protection from unauthorised people; Integrity requires protecting data from being altered; Availability requires ensuring access to data to authorised users when needed; Non-repudiation requires an assurance that authentic communication requests cannot be denied.
A chain of trust is established by validating each process and component of hardware/software from the base up to the final system, including the design, manufacture, and supply chain. A dependency graph (chain of trust) can be created by examining the component and services in which one trusted layer establishes the trust in the next by validating it and providing the core trusted functions on which it depends. Any security weakness at a lower level compromises the security of the higher levels dependent on it. This results in an untrusted base that compromises trust in the system. The roots of trust for a system are levels of trust origin – the root of the chains of trust. The roots would be the hardware or hosting environment, the \ac{OS} and any applications, libraries, and compilers. For building a system in secure environments, the roots may be the factories and supply chains for the hardware, the software design processes for the libraries, the location of manufacture, the supply chain and delivery. Additionally, the data collected by the testbed can be sensitive such as patient health and environmental data. Processing sensitive data using data analysis and machine-learning techniques~\cite{mendula2020interaction} makes the testbed a target for cyber-criminals~\cite{9612604} and adversarial machine-learning attacks~\cite{9653662}.

\begin{figure}
  \centering
  \fontsize{14}{12}\selectfont%
  \resizebox{0.8\columnwidth}{!}{%
    \input{Diagrams/exposed-services-thread-model.latex}
    }
  \caption[System Design: Threat Modelling: Local and remote threat models originate from the bottom up and up to bottom respectively]{Local and remote threat models originate from the bottom up and up to bottom respectively}
  \label{fig:threatvectors}
\end{figure}
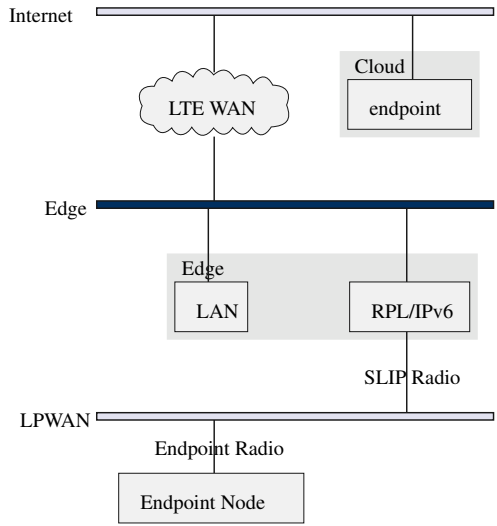
\subsubsection*{\textbf{Threat modelling - ``Identify threats, threat actors and determine risk acceptance''}}
Security of the testbed and the data collected is important. In projects that collect sensitive data, it is essential to understand the various threat actors, attacker models, and risks involved~\cite{Romn2013OnTF} that could compromise the security of the collected data or the testbed~\cite{KALLONIATIS2014759}. Creating a threat model is a crucial and challenging part of a research project and should be performed at the beginning of the project. It helps identify threats, attacks, vulnerabilities, and countermeasures that may affect the testbed infrastructure and its components. It can be performed in five significant threat modelling steps: defining security requirements, creating an infrastructure testbed diagram, identifying threats, mitigating threats, and validating that threats have been mitigated~\cite{Shevchenko2018, tmguide, mstm}. Threat modelling will enable testbed administrators and architects to communicate about the security design of the testbed, analyse those designs for potential security issues, and manage mitigations for security issues. 

An example of architectural consideration of threat models for our three-tier approach is presented in \autoref{fig:threatvectors}. Some key security questions arise, particularly regarding the edge and endpoint interaction. \ac{SLIP} bridges with the coordinator endpoint node require a multi-role endpoint node which requires separate firmware and networking behaviour for each node. The SLIP radio is the same hardware as other endpoint nodes but needs its firmware to be developed hand in hand with the edge device networking implementation to maximise security. The computing resources of the endpoint are minimal; therefore, communication with external devices must be tested with radio connectivity in full operation. 

\autoref{fig:threatvectors} also presents the concept of an edge network. Many challenges arise from the inability of the edge network to extend beyond a single computer (i.e. tunnel interface on a single RPi SBC). In this case, it is difficult to distinguish between the edge and the cloud and their interfaces. A strong firewall must be implemented and the network separation between \ac{LAN} and \ac{WAN} must be enforced at the edge site.

\subsubsection*{\textbf{Computation, hardware and physical security requirements}}

Based on the use cases and the required functionality, it is essential to determine the computation capabilities (memory, storage, \ac{CPU}) at each tier~\cite{Kurkovsky2017}. For example, cloud tier servers have high resources such as memory (8+ GB RAM), \ac{CPU} power (multiple cores), and network (Internet speed 50 + MB). Edge devices are \ac{SBC}s and have fewer resources (1-8 GB RAM, single or dual-core \ac{CPU}) than the cloud tier. On the other hand, endpoints are typically low in power consumption, memory (128KB-2MB of programmable flash and 20-512 KB of volatile RAM), and processing power (Arm Cortex-M \ac{MCU})~\cite{Traces2015}.

Furthermore, devices at each tier should provide hardware security features such as a cryptoprocessor (\ac{TPM}), the hardware-based root of trust that allows secure boot, secure firmware, secure credentials storage, and an encrypted file system. Secure boot prevents the loading of unauthorised software onto the device during the boot process; Secure firmware ensures that only authorised code (signed images) from the manufacturer is booted. Secure boot and firmware update capabilities ensure that the device does not run unauthorised or malicious code. Crypto-processor with a random number generator enables cryptographic functions such as encryption, decryption, and key generation for security purposes. However, generating random numbers in constrained embedded systems is a significant challenge due to the lack of resources and entropy. Modern endpoints provide a way to protect the integrity by providing a physically write-protect non-volatile memory with a mechanical switch. The end-user can switch to write-enable the memory for firmware update and then write-protect the device once the update is complete.

Furthermore, the physical security of the devices is essential, as they can contain confidential data such as \ac{PII}, log-in credentials and network information. An attacker who can gain physical access to devices can compromise and steal confidential information. Cloud tier servers hosted inside a secure perimeter (company offices) are physically more secure than the devices at the edge and endpoint tier deployed in the field (citizens' houses, bridges, streetlamps, or roadside). A determined attacker can reach the physically insecure edge and endpoint device and compromise its security. For example, an attacker having physical access to the edge device that contains an \ac{SBC} (e.g. RPi) can easily remove the \ac{SD} card and read its contents containing confidential information such as passwords and data. To provide another example, the AoT node (deployed on out-of-reach streetlamps) exposes a serial cable wrapped in a protected rubber cover connected to the \ac{UART} of the SBC. It can provide access to the device enabling the node's root access and allowing access to the filesystem and possibly confidential data. During the AoT project, it was found that it is essential to place edge and endpoint devices outside public reach (where possible) and protect them with spikes, locked cabinets, and tamper-proof casing.

However, once attackers have physical access to the edge and endpoint device, they can physically manipulate it to compromise them. The edge and endpoint tier devices have a large attack surface area, such as exposed copper vias and unused connectors, such as serial/Joint Test Action Group (JTAGs) used for debugging. An attacker can extract confidential data and embedded firmware code from the device using physical probing signals on the exposed interfaces. Most endpoint devices contain a sticker detailing the hardware components that can provide additional information to hackers. Devices with adequate physical and hardware security make it difficult for attackers to compromise them.

\subsubsection*{\textbf{Resilience (network, device, thermal, power and testbed)}}

Edge and endpoint nodes deployed in citizen houses or public spaces connect to the Internet and cloud via home broadband, Fibre, 4G, or Wi-Fi. The average downtime of broadband per year ranges from 25.4 to 168.9 hours in the UK~\cite{BristolBroadband}. Suppose the edge and endpoint device sends the endpoint data directly to the cloud tier without storing it locally. In this case, data will be lost due to lack of network connectivity~\cite{Schleich2012, Lundrigan2019, Jackson2017}. Furthermore, applications also suffer from latency problems~\cite{In1998} depending on the quality of the network. It is essential to have network resilience (multi-network such as Wi-Fi, 4G, LPWAN) built into the device to handle network loss and latency issues.

Furthermore, there can be scenarios where the edge node becomes unresponsive, does not connect to cloud services, and cannot be accessed using \ac{SSH}. In such cases, building resilience on edge devices is good. For example, AoT~\cite{Catlett2017} implemented a waggle manager to monitor the health of the SBC (temperature, current draw, digital heartbeat), enclosure internal temperature and humidity. It supports changing the boot medium from SD card to \ac{eMMC} and allows a hard and soft reset of onboard sensors. Rebooting the device often solves most problems~\cite{reboot_comp}. In such cases, a mechanism to reboot the device remotely is required. For example, if the edge device has multi-network connectivity (LoRaWAN, Sigfox, NB-IoT) and is not responding, the cloud tier can use LPWAN to send a downlink packet destined for that device, instructing it to reboot the system. NFC or magnetic devices can be used to cold-reboot the device without opening the enclosure (helpful for cold-rebooting publicly deployed devices)~\cite{Sotres2017}. If the devices are powered by \ac{PoE}, the ability to remotely turn the device on and off \ac{PoE} is preferable. The edge device should also be able to operate independently if cloud tier services are unavailable due to network issues~\cite{Catlett2017}.

Edge and endpoint devices generally run 24 $\times$ 7 and are usually deployed on citizens' premises or streetlamps. Suppose that a processor-intensive application is performed on the endpoint or edge, and the amount of processing power on the device is not regulated. The device can be damaged due to overheating. For example, in SPHERE houses, the Kinect camera that captures the activities in the kitchen runs 24 hours a day, processing the data. The camera becomes quite hot, reducing the device's lifespan. The edge and endpoint device should be able to self-regulate its temperature by performing \ac{CPU} throttling to reduce the temperature. For example, RPi performs \ac{CPU} throttling when the device temperature reaches 60-80 degrees~\cite{rpi_thermal}.

Another challenge is to provide electrical power to devices at the edge and endpoint tiers. Edge tier devices are usually powered by a mains or battery and must be safe from an electrical safety perspective. For example, AoT is powered and installed on the streetlamp with a 110/230V mains supply. An electrical hazard can occur should the device fall from the streetlamp or the transformer inside the device malfunction. The edge tier devices deployed on the streetlamp can be powered by \ac{PoE} to reduce electrical risks. Running on the battery limits the device's capabilities. Battery lifetimes typically range from a few hours to a few days. For example, SCK kits provide a \ac{USB} rechargeable battery that lasts for at least a day, depending on the sensing interval and the time to send the sensor data (after 30 seconds or 1 minute). Additionally, the use of solar panels can add resilience to power devices.

Additionally, a testbed can contain development, staging, and production environments. The testbed environment will often be compromised by an attacker creating a cyber security incident due to default credentials or misconfiguration~\cite{Keahey2020}. Once the testbed is compromised, it is essential to understand the affected components, as the attacker might have installed difficult-to-detect rootkits. It is prudent to recreate the entire testbed environment from scratch automatically. If done manually, the entire activity (setting up the VMs, configuring the applications, and ensuring that the end-to-end system is working) can take up to a week or more. To quickly recreate the testbed environment, it is essential to have version control~\cite{Kurkovsky2017}, continuous integration, delivery, and automation.

\subsubsection*{\textbf{Authentication, Authorisation, Certificate Authority (CA) and secure storage of secrets}}
Testbeds consist of multiple devices and numerous applications on the cloud or at the edge for data storage, analysis and visualisation and have multiple users/administrators accessing those applications and devices. Devices and applications should have proper authentication and authorisation, allowing trusted users to access services~\cite{soldatos2015openiot}. Authentication requires digital certificates or credentials to validate the identity of devices and users. Authorisation requires that only trusted nodes and users should be able to gain network access to the testbed. As the testbed also hosts different services (such as web servers, WebSockets, and authentication servers), it is essential to have a CA in the testbed that can be used to create public-private keys and sign certificates. Different users and devices can trust the CA to secure data transmission.
Further, the testbed will need to protect stored cryptographic material. The encryption keys (public/private and symmetric) and credentials are usually hardcoded in the code or stored in files. To protect the credentials from hard coding and unsecured storage, they must be stored securely using a hardware security module or key management solutions.

\subsubsection*{\textbf{Exposed services and security updates on the endpoint, edge, and cloud}}

Devices in each tier run multiple services (e.g. \ac{SSH}, web servers) and are often insecure with weak authentication mechanisms. These mechanisms include using default passwords, running a vulnerable version, using old encryption methods, and misconfigured applications~\cite{Keahey2020, Mydlarz2019}. The services exposed on the cloud, edge, and endpoint devices depend entirely on the project's requirements. Additionally, the greater the number of services, the greater the attack surface area for the attackers and the possibility of compromise.
For example, the cloud can expose port 1194 (\ac{UDP}) and \ac{TCP} port 443 to provide \ac{VPN} connectivity. The Grafana server (data visualisation) exposes port 3000. An edge node might expose port 1883 to allow communication with endpoint devices using \ac{MQTT}. The endpoints can also run a Web server~\cite{contiki_webserver}. As endpoints are resource-constrained, there is a possibility that they might be running a vulnerable version of the web server software.

There have been instances where attackers have compromised insecure services running at the cloud/edge tier. For example, an attacker compromised a cloud server providing authentication (Keycloak instance) running with default credentials and used the server for crypto-mining~\cite{Keahey2020}. Alternatively, an internal attacker can connect to the insecure MQTT service running on the edge device and subscribe to the topics to collect the published data. Furthermore, a vulnerable application deployed on the cloud/edge poses a security risk.

However, such services and systems must be made secure by default. It is essential to ensure that there are no default passwords and that the \ac{OS}, applications and firmware are configured securely and up to date. If the infrastructure contains many devices kept remotely (citizens houses, streetlamps), upgrading software/firmware is often challenging. Software updates should have rollback functionality, so the device will return to its previous state even if the update process goes wrong. Upgrading software is comparably easier than upgrading firmware. A poor firmware update mechanism can leave the device unusable when an update fails.

For endpoints, it is recommended to have \ac{OTA} functionality to allow remote upgrade and configuration for long-term deployments in urban environments~\cite{Dehwah2015, Elsts, Barrenetxea2008}. The inability to upgrade or configure the firmware remotely means that the code/firmware must be perfect and thoroughly tested, and no new requirements can be applied. For example, the Cotham Hill Pedestrianisation Programme wanted to measure noise pollution due to pedestrianisation. However, the deployed SCK kits took sensor readings at 60-second intervals (by default) and did not capture noise pollution correctly due to the 60-second gap. The only way to reduce the reading interval was to revisit the citizen's houses and configure the settings resulting in disturbing the citizens. Remote management of the technology will minimise disruption for the participants.

\subsubsection*{\textbf{Data storage, reduction, access, integration and visualisation}}

Research projects require data storage, analysis, and visualisation. Data must be encrypted in transit and rest at all tiers. Research projects often go through different data protection and research ethics, defining data collection and usage. The data owner's responsibility is to ensure data validity, quality, secure storage, access and maintenance, replication, processing, backup, and deletion policy. Having clear information and policies helps to ensure user privacy~\cite{8896155}. Policies should include what participant data will be acquired, where it will be stored, and how long it will be stored. User data should be deleted once the duration of data consent is over. However, Post Docs/Ph.D (staff joining and leaving) often manage research projects, and it becomes challenging to ensure data deletion. For example, in university-managed research projects, access to the data is usually restricted to university premises (IT services managed machines) and provided via jump host machines via different credentials, and might require hopping through multiple networks. The difficulty in accessing the data makes it challenging for the data analysis activity, resulting in researchers copying and processing the data locally, which may break user privacy and data agreements.

Further, sensitive data can attract attackers. It is ideal to identify potentially sensitive information in the collected data at the endpoint/edge tier and eliminate or limit its collection~\cite{KumarR2019, Staudemeyer2019}. Data reduction and compression methods, such as sending preprocessed data to the edge/cloud tier rather than raw data~\cite{Folea2020}, can also help reduce data bandwidth and power consumption. For example, an edge tier device that measures the number of cars parked using image recognition should send only the count rather than the images to the cloud~\cite{KumarR2019}. Another example would be when an endpoint only transmits the reading to the edge device when a significant change is detected to improve the energy efficiency of battery-powered endpoints~\cite{Elsts2018}. Data compression and reduction should maintain the initial data requirements required for the research project's objective.

It is a good practice to store all raw data for historical and future references~\cite{beaudin2004lessons}. As users frequently access the collected data of the last few days, it is a good practice to separate current and historical data for better application performance~\cite{Valera2018}. For example, 3E houses executed SQL queries on the sensor data recorded. Over time, the query response time changed from < 1s to > 8s, resulting in an unresponsive display leaving citizens less engaged~\cite{Porto2013}.

From a data integration perspective, the platform should be able to integrate data streams from multiple heterogeneous data sources~\cite{6924615, villanueva2013civitas, 6815176,galache2014clout}. Using similar data formats will allow better data interoperability~\cite{Chatterjee2018, In1998, girtelschmid2013big}. Further, the testbed should provide an open \ac{API} for the end-users/developers to access the data and build applications on top of that~\cite{In1998, Cheng2015, Elsts2018, Chowdhury2018, 6525605}. Furthermore, data transfer from the endpoint to the edge to the cloud should be reliable with minimal data loss~\cite{Lundrigan2019, Basnyat2020}. During the AoT and Cotham Hill Pedestrianisation project, it was found that providing flexible data query capabilities for users (such as extracting specific periods or a subset of measurements/nodes) is essential. Such capabilities allow the user to monitor conditions over a particular period, such as an ongoing event (e.g.\ a festival, severe storm, or emergency), and stream data to specific stakeholders (city-council/car-parking and others). Data should also be visualised for stakeholders using different methods (maps, line/bar charts, dashboards and others)~\cite{soldatos2015openiot}.

\subsubsection*{\textbf{Technology compatibility, Device naming conventions and Time synchronisation}}
The testbed comprises multiple components, including hardware, software and \ac{OS}, to support various services such as data storage, analysis, visualisation, authentication, and authorisation. In addition, there could be different hardware platforms such as amd64, armhf (32 bits), arm64 architecture \ac{CPU}s, \ac{GPU}s, and \ac{TEE}. It is vital to support standard libraries, packages (for researchers to deploy their applications on the device), and control interfaces (USB, I2C, SPI, serial) to add new hardware modules with standard network technologies (Wi-Fi, wired, Bluetooth)~\cite{Catlett2017, Oyedele2021}. Creating an interoperability matrix that captures the different versions of software and the OS is important. For example, Debian 11 switched to cgroup v2, which broke some applications (docker monitor)~\cite{Internet:broken_docker}.

The platform can contain hundreds of thousands of endpoint and edge devices. It is essential to have a good naming convention for devices at each tier to identify them uniquely and the data generated from the devices~\cite{Cheng2015, Dehwah2015}. Also, all devices in each level (cloud, edge and endpoint) must be synchronised in time for data integrity and audit log purposes~\cite{Elsts2017, Barrenetxea2008}.

Requirement analysis helps to understand the research project's aims and objectives. System design helps to understand how the set of requirements can be achieved. Once a higher-level system design is defined, the testbed architect can start implementing the testbed architecture, functional model~\cite{Gea2013}, and how devices at the endpoint, edge, and cloud tier will be managed, provisioned, and communicate with each other~\cite{Chowdhury2018}.


\subsection{Implementation}
The implementation phases bring challenges such as provisioning devices, ensuring secure network connectivity, credential management, application deployment, and compatibility between different hardware architectures (armhf, arm64, amd64), hardware and software accounting and monitoring. The challenges of the integration phase include ensuring that the platform is scalable, modular, extensible, adaptive, and reproducible and supports heterogeneous devices, proprietary software, and different standards.

\subsubsection*{\textbf{Provisioning the cloud, edge and endpoint devices}}
Provisioning the cloud tier requires the installation and configuration of \ac{VM}s on the on-premises hosted hypervisor (Hyper-V, Proxmox, OpenStack) or cloud hosting providers (AWS, Azure). The number of VMs depends on the services required to support the edge and endpoint tier and usually ranges from one to ten. Installing and configuring a \ac{VM} is a tedious task and requires installing \ac{OS} applications, configuring them securely, and configuring hardware allocation (e.g. RAM, \ac{CPU}s, GPU passthrough). Most research projects currently provision the servers manually or using a bash script. The bash script installs the necessary packages and configures them with security. Those images can be packaged to support different hypervisor environments without requiring changes to the provisioning scripts and source code. Such platform-independent virtual machine image creation tools are Yocto and Packer.

Provisioning edge tier devices (Intel NUC or \ac{SBC}) involves installing an OS on the \ac{SD} card/\ac{HDD}/\ac{eMMC}, with configured software packages, and ensuring stable and secure connectivity to the cloud tier. The number of edge devices depends on the sample size of the case study, such as the number of houses or streetlamps, and can range from one to hundreds.
One way to provision edge devices is to create a base kernel image containing the installed OS and applications and flash it to the edge devices. Adding the Linux kernel headers in the base image is essential because future application installations might require building a kernel module (e.g. wireguard). Otherwise, the base image needs to be created and flashed again. For any further changes, the administrator logs in to the device using the \ac{SSH}/serial console and configures it according to the requirements.
Creating a base image and flashing it on multiple edge devices comes with security and administration challenges. The security challenge is that the credentials and other settings, such as Wi-Fi SSID, hostname on all the edge devices, will be the same until changed. If one of the edge devices is compromised and the attacker obtains the credentials, they can compromise all the edge devices by performing the lateral movement. The administration challenge is to log into the machine and make changes after flashing the base image. For example, deploying the edge device in the citizen's home could require changing parameters such as house number identification, Wi-Fi credentials, and IP address settings. Additionally, suppose that the device is deployed on citizens' premises during pandemic outbreaks. In that case, minimising the time spent configuring the device is essential.

Endpoint tier devices are usually resource-constrained devices, such as \ac{SCK}~\cite{sck_website}, Luftdaten~\cite{luftdaten_website}, SensorTag~\cite{sensortag_website}, and Smart Plugs~\cite{smartplug_website}. Endpoints are usually connected to the smart home platform or the edge device. The provisioning of endpoint devices depends on the capabilities of the device and the communication medium between the endpoint, edge, and cloud. It mainly includes configurations such as setting up the connectivity (using Wi-Fi/ZigBee/802.15.4), the MQTT server address to publish sensor data, and the time at the endpoint using \ac{NTP}. Moreover, standards such as \ac{LWM2M}~\cite{AllianceOpenMobile2012} have been developed to manage endpoints securely and in a mannered function. LWM2M provides device management capabilities (remote provisioning of security credentials, firmware updates, and connectivity management) and service establishment capabilities (sensor readings, remote actuation, and endpoint device configuration). Various papers~\cite{Fafoutis2017, Elsts, Langendoen2006, Barrenetxea2008} have provided lessons learnt from experience by deploying battery-powered devices in the endpoint tier communicating over IEEE 802.15.4.

Endpoints could also be configured dynamically or bootstrapped by the device on the edge/cloud tier by providing configurations such as which endpoints are allowed to join the network, the encryption keys to encrypt the data, and the network address/port number of destination, and other settings. Additionally, communication between the endpoint and the edge must be encrypted. For example, if the endpoint connects to the edge via 802.15.4, the edge device requires a border router to communicate. If the endpoint connects to the edge via Wi-Fi, Wi-Fi encryption (WPA2) encrypts the data over the air. For example, the SPHERE~\cite{Elsts2017} project deployed multiple endpoints connected using 802.15.4 in around 100 houses in Bristol and used one hard-coded encryption key per house to encrypt data over the air. They used media access control (MAC) address filtering to prevent external devices from joining the IEEE 802.15.4 Time Slotted Channel Hopping (TSCH) network.
\subsubsection*{\textbf{Endpoint-Edge-Cloud Connectivity}}

From the communication perspective between devices at each tier, it is essential to use encrypted protocols for communication from endpoint to edge to cloud tier~\cite{Elsts2017, Fafoutis2017}. Secure transmission protects against packet sniffing, man-in-the-middle attacks, replay attacks~\cite{9810983}, and unauthorised attempts to communicate with the node.

\begin{figure}
   \fontsize{14}{12}\selectfont%
  \resizebox{\columnwidth}{!}{%
    \input{Diagrams/3tier-wsn-data-path.latex}
    }
  \caption[Implementation: Endpoint-Edge-Cloud Connectivity: Connectivity points between the three tiers for a WSN use case]{Connectivity points between the three tiers for a WSN use case}
  \label{fig:iotcloud}
\end{figure}
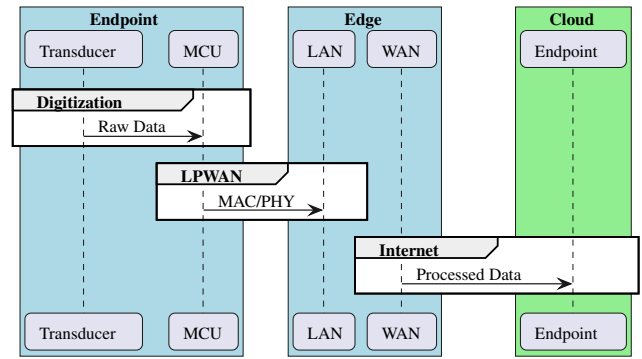

The servers that host the cloud tier must provide services to edge tier devices and expose them to IP addresses and ports. Services could range from \ac{HTTP}, HTTPS, WebSockets, \ac{LDAP}, \ac{VPN}, and others and may require different ports exposed to the Internet. Testbed administrators prefer to reduce the number of ports exposed to the Internet to reduce the attack surface area, which is better from a security perspective. An example of a \ac{WSN} implementation providing the connectivity points between the three tiers is presented in \autoref{fig:iotcloud}.
Both sensor \ac{LPWAN} nodes and cloud addressable \ac{URL} or IP can be considered endpoints. The challenge for the edge device is to distinguish between the two directions of communication. Routing tables for packet forwarding for \ac{LAN} and \ac{WAN} and also the \ac{SLIP} bridge create complexity and are challenging to design, implement, and secure.

\textbf{Edge to Cloud Connectivity:}
There are three ways to expose services hosted on the cloud tier. Firstly, by opening the ports on the cloud tier firewall. However, opening multiple ports on the firewall increases the attacker's surface area and is not preferred~\cite{LeBlond2012}. Second, connect the device through \ac{DMZ} to the cloud using a \ac{VPN}~\cite{Mydlarz2019}. However, in the case of a cyber-incident where an attacker compromises one edge tier device, they can explore and enumerate the internal network for vulnerabilities (depending on routing configuration and if the network is flat at the data-link layer). The third is to use a \ac{SDP} that runs a client on the device using the authentication process. SDP defines a policy to determine who gets access to what resources and distributes access to internal applications based on a user's identity. It makes the application infrastructure invisible to the Internet, evades network-based attacks (DDoS~\cite{9810983}, ransomware, malware, server scanning) and reduces the security risk. However, enterprise organisations often use SDP, which might be overkill for a research testbed. Furthermore, if the devices at the edge and cloud tier are in the same network connected over ethernet or Wi-Fi for demonstration purposes, edge and cloud tier devices will be in a trusted private network; VPN or firewall might not be required.

The typical way to connect edge devices to the cloud network is through a VPN. For example, if there are 50 edge devices in different houses or streetlamps, it is good to generate 50 unique credentials from a security perspective. However, more manual/scripted effort is required to create credentials and provision them to nodes. For example, the REPLICATE project used OpenVPN to provide secure connectivity and issued certificates through a CA. The administrator generated 150 credentials and stored them on a USB stick with 150 folders for each house. The deployment team (DT) was responsible for visiting a particular home and installing and provisioning the edge and endpoint devices. They executed the bash script on the edge tier device that installs the certificate for that house and provides secure connectivity to the cloud tier.

\textbf{Endpoint to Edge Connectivity:}
Endpoints are usually connected to the edge/cloud using mesh networks and LPWAN technologies. The choice of network technology depends on connectivity requirements such as range, bandwidth, power, interoperability, security, and reliability~\cite{Chatterjee2018}.


However, there are challenges when multiple endpoint devices communicate over various channels in an urban environment. An urban environment can have numerous networks such as cellular, LPWAN, mesh, and others. In a real-world deployment, connectivity between multiple devices in the vicinity of each other depends on external interference, frequency-selective multipath fading, and dynamics in the environment. The dynamics of the environment can include the number of people, the movement of people, the Wi-Fi traffic, the rooms, the layout, and the type of building materials used~\cite{Elsts2018, Huang2017}. A house deployment might initially function until further technology is deployed into a neighbouring house, causing disruptions due to radio interference. External interference can occur when a different technology or a deployment of the same technology operates within the same radio range (IEEE 802.11 Wi-Fi interferes with IEEE 802.15.4 at 2.4 GHz)~\cite{Brun-Laguna2018, Traces2015, 5779219}. Furthermore, in an 802.15.4 network, the mobility and activity of an endpoint can affect the throughput and data on the mesh infrastructure.

\begin{figure}[t!]
	\centering
	\includegraphics[width=\linewidth]{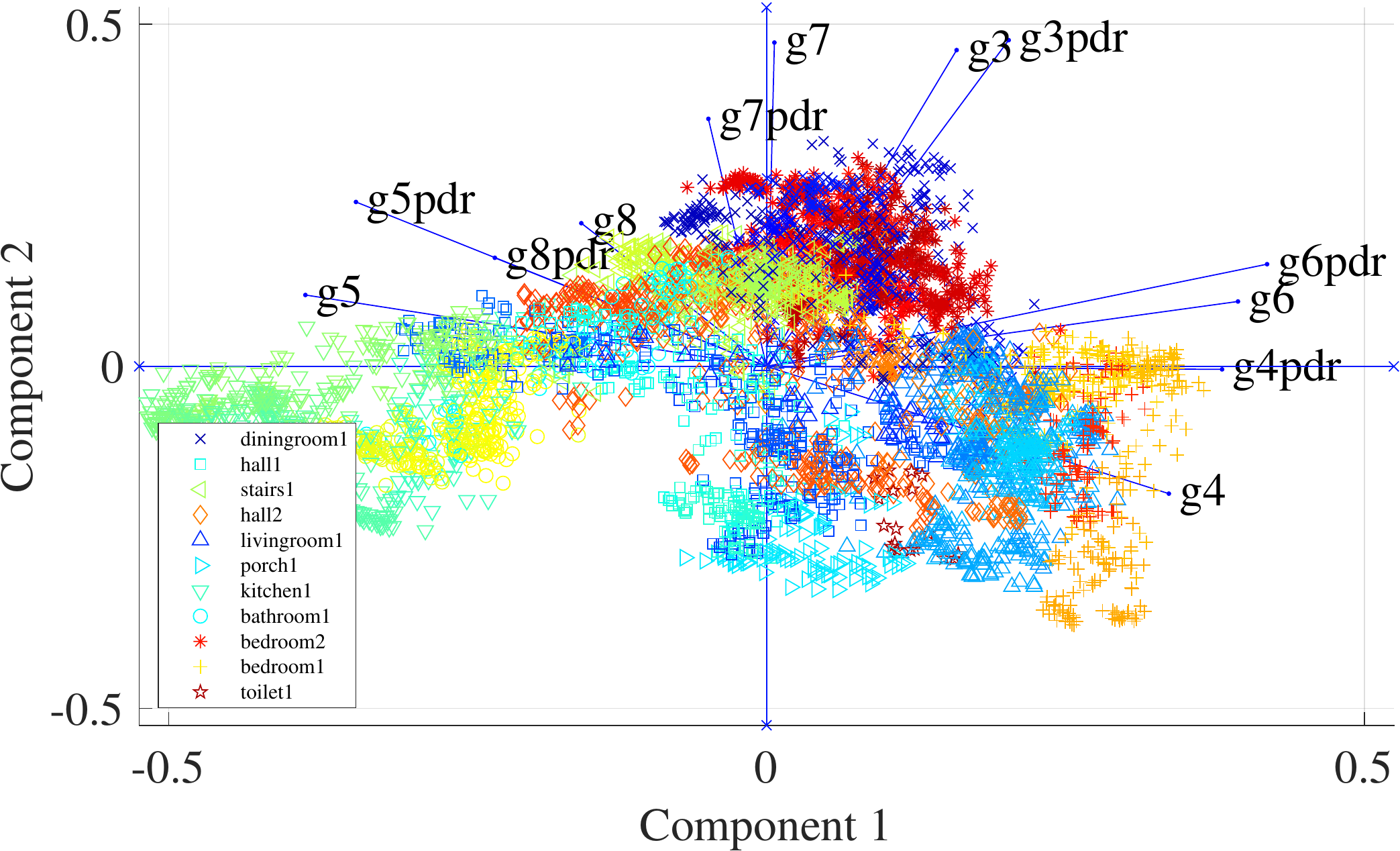}
	\caption[Implementation: Endpoint-Edge-Cloud Connectivity: Mobility of BLE tags in a house, the association of the PDR and signal strength for eight listening gateways]{Mobility of BLE tags in a house, the association of the PDR and signal strength for eight listening gateways}
	\label{fig:pca_pdr}
\end{figure}
\autoref{fig:pca_pdr} presents the \ac{PDR} calculated from packet sequence reconstruction for individual receivers in a home environment. The strength of the received signal and the packet loss patterns show the effect of mobility between rooms in the residential environment and the effect on \ac{PDR}. The \ac{PDR} is affected by the increased bandwidth requirements on the forwarding gateways when many packets are generated locally that require forwarding. In \autoref{fig:pca_pdr}, four tags that require a fixed uplink bandwidth generated enough packets to saturate the uplink capacity allocated to the mesh network. In particular, gateway 8 is sharing uplink bandwidth with gateway 5, which is visible from the alignment of the two \ac{PCA} components of \ac{PDR} (g8pdr and g5pdr). In other words, gateway 8 uses gateway 5 in a mesh network topology to forward its traffic in the network. Since the available bandwidth is limited, there is a lot of packet loss in the data originating from gateway 8, making the PCA component \textbf{g8} the least significant in the overall entropy. 
The \ac{PDR}, network usage, and packet loss have a dynamic nature in a dynamic environment~\cite{Langendoen2006}. For example, SPHERE has deployed a mix of network technologies such as 802.15.4 400 MHz, BLE channels 37, 38, 39, and 16 channels of 802.15.4, 5GHz Wi-Fi, and a router with an Ethernet interface.
BLE packets were generated on the advertisement channels 37, 38, and 39 with an interval defined by the BLE 4.2 standard at about every 200 ms. The specification allows only a fixed interval with increments of 0.625 ms with a random delay of 0 ms to 10 ms. These packets are scanned from receivers that scan on one of the three channels at any particular time and rotate across those channels many times every second. Those packets are encapsulated in CoAP messages, which are forwarded to the 802.15.4e gateway from these intermediate receivers using a fixed uplink time-slotted schedule. The gateway uses a bridge to bring CoAP messages to a compute host using Contiki-NG \cite{oikonomou2022contiki}.
Link quality is an important metric when connecting endpoint devices to the edge/cloud. When the security of the communication channel depends on the \ac{RF} channel, if an attacker gets physical access to the device or sniffs the network, they can learn the procedure for joining the network, such as the exchange of network keys.
In particular, in IEEE 802.15.4, in the minimal implementation, the pattern of connecting a node to a network uses a fixed channel \cite{dujovne20146tisch}. Information for the particular network in its formation~\cite{contiki_webserver} can be inferred by sniffing those 10 ms timeslots where routing is established~\cite{8998289}.

\subsubsection*{\textbf{Credential Management}}
After provisioning, the edge and cloud devices must be maintained and accessed occasionally. One of the ways to access the device is by \ac{SSH} using authentication mechanisms or credentials such as a username, password, or digital certificates~\cite{Mydlarz2019, Belli2015}. The device can authenticate the user by storing the credential on the device or authenticating through a central server and storing it locally for a specific time. Using passwords is not recommended, as it allows the attacker to brute-force the username and password. Furthermore, when the password is sent to the device for authentication, it can be compromised by \ac{MITM} attacks~\cite{LeBlond2012}. One preferred way of providing access is to store the administrator's public SSH keys\footnote{SSH has public and private keys, the public key is stored on the device, and the private key is kept with the user requiring device access.}~\cite{Gran2014} in each of the devices. However, storing public SSH keys on the device is risky as if one of the private \ac{SSH} keys is compromised, access to all edge devices may be compromised. In addition to using SSH, administrators also use remote management tools such as TeamViewer/AnyDesk to update scripts or perform functionality that requires \ac{GUI}. However, recently attackers compromised Florida City's water supply using remote access software (TeamViewer), which allowed staff to share screens and troubleshoot IT issues~\cite{teamviewer_water} by exploring systems from the Shodan search engine and outdated passwords.

\subsubsection*{\textbf{Application deployment and compatibility on different systems}}
Research projects involve multiple researchers developing different applications (Python/R programs)~\cite{soldatos2015openiot} that need to be deployed on the edge device with different architectures (arm64, amd64, armhf). Researchers need to access edge device hardware (sensors, cameras, GPU) for edge processing and cloud resources for data analysis. Initially, developers work on sample data and develop applications that work fine on their machines. However, applications must be deployed on the edge and in the cloud to access real-world data. Deploying custom applications often requires installing library dependencies (e.g.\ pandas, scikit-learn) and may require administrative privileges, often resulting in the application not working correctly on the edge/cloud platforms. 

The above results in scenarios where developers say, ``It works on my machine!'' resulting in numerous meetings and debugging of applications to determine the root cause of the problem. Python and Linux distributions have a lot of inter-component dependencies embedded into them. It is crucial to monitor those interdependencies and evaluate any security updates against those dependencies. Tools are being explored in the literature to evaluate those dependencies~\cite{DelSole2021,5767626} and provide early warning when changes lead to incompatibilities.

Additionally, the project must always store the data collected on designated machines to comply with data protection laws and user privacy. Many applications need access to a graphics card or more memory to process the data. This requires moving the data to a more computationally capable machine, which becomes challenging due to data management guidelines. Due to data management guidelines, application incompatibility often results in either no or delayed application execution on the whole dataset. The application code also needs to be consistently deployed on devices; one of the ways it is maintained is by using a remote git repository cloned on the device remotely updated as a batch process~\cite{Lundrigan2019}.

\subsubsection*{\textbf{Accounting and Monitoring}}
The testbed can contain tens, hundreds, or thousands of devices on the cloud, edge and endpoint tiers. It is crucial to maintain an inventory of the number of devices at each tier, with their hardware and software details (make and model, OS versions, installed applications, and their version)~\cite{beaudin2004lessons}. The OS and application version can be used to actively monitor the \ac{NVD} database to detect vulnerabilities and patch the system proactively.
Additionally, audit logs with synced timestamps should be collected to a central server and enabled to ensure forensic investigation during cyber-security incidents. Also, it is essential to maintain the details of who (i.e., which user) has logged into which machine and performed what activities for auditing purposes. However, it can depend on the remote management software's licence (free version/enterprise edition).

The infrastructure deployed for data collection requires that all hardware/software be working as expected and usable by researchers~\cite{Gran2014}. In addition, all endpoints must be connected to the edge, which should be connected to the cloud tier. If not, any loss of network connectivity can result in data loss. The monitoring infrastructure is essential to ensure this~\cite{Keahey2019, Elsts2017, Mydlarz2019, Elsts2018, Hnat2011, Sotres2017}. Monitoring includes detecting whether devices are reachable and sending regular data. Monitoring also includes checking infrastructure components (such as web servers, adequate disk space, and system overload)~\cite{SYED201711}. The monitoring infrastructure should include an effective alert mechanism (email, slack, text messages). From the endpoints deployed through 802.15.4, it is good to have statistics about energy (battery), network (number of data/control packets, acknowledged packets), neighbourhood statistics (list of neighbour nodes and the link quality), per-channel per-neighbour packet reception rates, TSCH time synchronisation performance, background noise \ac{RSSI} levels, stack usage, and others~\cite{Elsts2017, Barrenetxea2008}. For example, SPHERE~\cite{Elsts2018} monitored the status (reachability) of the deployed endpoints by regularly polling various devices within the home network based on Nagios.

\subsection{Integration Testing}
\label{subsec:integration}

After the system design and implementation of the testbed, it is vital to perform integration and testing at regular intervals, such as ensuring that interlinked components are working correctly; the platform is scalable, modular, and extensible; integration of heterogeneous devices, proprietary software, and different standards; ensuring endpoint and edge provide good ruggedisation; ensuring testbed adaptiveness and replicability.

\subsubsection*{\textbf{Interlinked components dependency}}

Data gathering research projects have multiple interdependent components and interfaces installed on devices to ensure data transfer from the endpoint to the cloud. A component is the system's part/block (hardware/software). On the contrary, an interface is a part that connects two or more other components to pass information from one to another~\cite{4116787}. It is the mechanism through which the components of the block communicate. For example, a web server is a component, and the HTTP/WebSockets (method of communication) will be the interface. The glueing of software components requires considerable effort and in-depth knowledge of the components~\cite{Langendoen2006}.
The data generated by the endpoint follow a pipeline and travel through multiple interconnected components to the cloud. Each component expects the data to be in a specific format or size. Often, a component might fail to pass the data to the next component in the desired form, failing the whole pipeline~\cite{Oyedele2021}. For example, an endpoint sends the data (such as temperature readings) through MQTT in \ac{JSON} format to the edge device for processing and storage in an InfluxDB database. The edge device can run a Python script to check if the temperature is above a threshold and notify the cloud tier. There could be multiple points of failure in this example, such as issues in MQTT, wrong JSON format, InfluxDB server not running, python script error, and others.

An administrator often needs to buy several devices with different components and interfaces for a research project. They need to learn how the devices work, test them, ensure that the data can be fetched in a limited amount of time in a lab environment in a specific setting, and finally deploy them in the wild~\cite{Dehwah2015, Basnyat2020}. For example, research projects that involve energy monitoring deploy multiple devices such as smart plugs~\cite{TPLink_SmartPlug}, Tesla powerwall~\cite{Tesla_PowerWall}, OpenEnergyMonitoring~\cite{openenergymonitor}. When deployed in the real world, there is a probability that a system component will not work as expected due to hardware or software failure~\cite{Valera2018}. Debugging and finding the misbehaving piece takes considerable time and is challenging~\cite{Conti2014, Ekedahl2018, Dehwah2015, Kurkovsky2017}. It requires detailed logs of different system components with timestamps, understanding what triggered the logs, and ensuring that the devices generate log messages representing various failures.

Therefore, performing regular automated integration and end-to-end testing is essential to prevent such failures~\cite{Webb2004}. Additionally, components and their functionality must be well defined and have robustness and resilience built in, saving time for system administrators~\cite{Elsts2017, Elsts2018, Chatterjee2018}. It also helps minimise the number and duration of visits to the citizen's residence to repair the system~\cite{Lundrigan2019}. The maintainability of the infrastructure and the consistency of the interfaces between all different components~\cite{Nussbaum2017} (such as \ac{COTS} of hardware/software) can help with the resilience of the infrastructure.

\subsubsection*{\textbf{Scalability, modularity and extensibility}}
Research projects require the deployment of endpoints in multiple locations. The testbed platform is easy to manage when small and consists of only a house/streetlamp in one place. However, running a scalable trial that is supposed to scale to 100-200 houses/location becomes challenging. The system must be able to scale to tens to thousands and tens of thousands of homes/streetlamps in a reliable manner~\cite{Valera2018, Elsts2018, Basnyat2020, bellavista2019survey}. In addition, software and hardware development occurs rapidly and can quickly become obsolete. The hardware and software components of the test bed must be designed with modularity and extensibility in mind to adapt to ever-evolving technology~\cite{Valera2018}. Hardware and software at the cloud/edge tier can be modular and extensible (for example, replacing the SBC at the edge with a newer, more powerful SBC)~\cite{Chowdhury2018}. However, modularity, extensibility, and future-proofing at the endpoint tier is challenging because it is difficult to predict the exact requirements of future deployments and the electronics market progresses quickly. As a rule of thumb, testbed designers should follow the \ac{KISS} principle~\cite{Chatterjee2018}.

\subsubsection*{\textbf{Heterogeneous devices, proprietary software, and different standards}}
Projects can have different devices on edge and endpoints generating various types of data and formats~\cite{KumarR2019, Gea2013, Hnat2011, Basnyat2020, 6924615, 8286847}. For example, edge tier devices can have SBCs (GrapeBoard, RPi, Coral boards, Intel NUC). Endpoints tier devices can have different devices such as Nordic Semiconductor nRF5340-DK2, Texas Instruments Launchpad (LAUNCHXL-CC2650/CC1310/CC1350), TI CC2650 SensorTag. The testbed requires the devices to be securely configured and connected to the network.
In addition, the endpoints used to collect data can run open-source or proprietary software~\cite{Gea2013, Hnat2011, Basnyat2020}. In the case of proprietary, they may not provide an open source script to take the sensor data and may have a \ac{GUI} to download the data or allow it to be sent only to the endpoint manufacturer website. In such cases, the administrator must figure out how to extract the data from the proprietary device or the manufacturer's website. Some proprietary technology may not be designed or evaluated for cybersecurity purposes. In addition, it is always difficult to evaluate and secure different network connectivity (802.15.4, BLE) in IP networks.

As there may be different devices from different vendors on the testbed, they can be running on various standards and formats (sending data over MQTT, HTTP, WebSocket, proprietary protocol), resulting in a lack of interoperability between sensors~\cite{Gea2013, Latre2016, Palattella2016, Ekedahl2018, In1998, wahle2012openmtc}. It is vital to use widely open standards and possibly the same standard and format to help reduce learning times for research personnel~\cite{Elsts2017, Chowdhury2018}.

\subsubsection*{\textbf{Ruggedization}}
Ruggedisation is essential when deploying devices in citizen houses or outside on streetlamps. For example, any edge device installed indoors/outdoors requires specific \ac{IPR} and electrical testing~\cite{Barrenetxea2008}. It must be packaged in a form that can be securely mounted~\cite{Catlett2017, Sotres2017} and still easily open if a battery or component change is required. \ac{IPR} define levels of sealing effectiveness of the electrical enclosure sealing against foreign body intrusion (i.e., dust) and moisture. From the electrical safety perspective, it is crucial to have a \ac{CE} rating (for EU/UK) or country-specific certification rating on the endpoint and edge device. The certification mark ensures that the manufacturer has verified that the products have met country-specific safety, health, or environmental requirements. For example, \ac{BUO} had difficulty installing \ac{AoT} nodes in streetlamps and on the university campus because the nodes did not have \ac{CE} ratings (the electrical safety certification of the USA is different from the UK).
Additionally, when designing enclosures for devices that contain sensors (such as air quality), it is essential that the airflow is optimal and allows the proper functioning of the sensors on board. The enclosure should protect the electronics from moisture and insects~\cite{Catlett2017}. It might be a good idea to place the sensors in a Stevenson radiation shield\footnote{shield instruments against precipitation and direct heat radiation from outside sources while still allowing air to circulate freely around them} separate from the sealed waterproof electronic enclosure. Furthermore, it is recommended to identify a suitable enclosure first (accepted and visually aesthetics) and then fit the edge and endpoint device in it with minimal modification. Designing a custom casing is often challenging and more expensive than modifying a readily available casing~\cite{Fafoutis2017}. During the Cotham Hill Pedestrianisation project, it was found that designing a 3D-printed enclosure, models, printing it, and post-processing the 3D print (cleaning up the support materials) is challenging and time-consuming.

\subsubsection*{\textbf{Testbed adaptiveness and replicability}}
The testbed must be adaptive to the project requirements or the community demand. For example, change in hardware requirements (such as a powerful graphics card, more RAM, hard disk space, or low-power processors) or human-interaction interfaces (ways to visualise/process data). Also, supporting as many users as possible depends on two factors: cost of users, experiments, and adapting the testbed to the needs of different communities~\cite{Keahey2020, Benzel2007}. Also, the testbed should be reproducible using open-source software and automation, allowing implementation of the testbed by other administrators using applicable documentation (e.g. wikis) and other supporting materials.


\subsection{Operational Testing}
The next step is to develop a prototype testbed in a laboratory and a small-scale real-world environment before large-scale deployment in the wild~\cite{Chowdhury2018, Jackson2017}.
\subsubsection*{\textbf{Time resource allocation}}
The concept of time as a resource available to the testbed can be interpreted as a \ac{CPU} processing time at both the edge and the endpoint. Furthermore, this can be associated with radio utilisation time at the co-coordinating endpoint connected to the edge or other edge nodes. The available time is governed by the data rate related to the sensor sampling frequency and resolution. Monitoring tools enable observations such as \ac{CPU} time use and radio usage, which is essential when scaling the testbed. 
To give some real-world perspective, a byte of data, when transmitted, is serialised into eight bits of 0's and 1's and sent over a medium such as wires or radio. Communication protocols are responsible for encoding/decoding the bytes and bit streams and depend on the medium's capacity in bits per second.
This can create an interesting paradigm between radio use and environmental monitoring. Almost all analogue-to-digital converters support Layer 2 access control allowing many sensors to be connected to inexpensive \ac{SoC} micro-controllers. This reduces the cost of the \ac{PCB} design by reducing the number of wire traces and complexity. Similarly, the radios, where the MAC layer controls access to the radio medium. In both cases, consideration of time allocation applies. 

\subsubsection*{\textbf{Lab deployment}}
The testbed will contain multiple heterogeneous devices at each tier. Each device would have different interfaces, components, applications, and services running. It is essential to ensure that the system is working as a whole~\cite{Auer2020} and securely sending the data from the endpoint to the cloud with analysis and visualisation satisfying project requirements. The platform must be deployed in a laboratory environment before being deployed on a large scale. It helps to face the challenges early on and test any new software/application internally on the testbed rather than pushing it directly into production.

Assignment of a provisioning budget is essential for setting up a lab testbed, buying various spare devices and components, and conducting deployment site visits. Based on the budget, project scope, and the number of researchers working, it might be good to have more than one lab testbed (dev1, dev2). Multiple lab testbeds help keep work in progress, even if one testbed has broken down because of a misconfiguration or software/hardware failure. Additionally, the laboratory testbed must be set up and running as early as possible in the project to test the different devices, components, software updates and applications to ensure the final real-world deployment is completed on time. Although only sometimes possible, the testbed should be as close as possible to real environmental conditions. For example, the Living Lab project first deployed electrochemical air quality sensors using laboratory-based wall sockets; however, electromagnetic interference from the power supply caused interference in the sensors, affecting the readings when deployed in the field~\cite{Jackson2017}.

\subsubsection*{\textbf{Small scale real-world deployment}}

\begin{figure}
    \centering
    \includegraphics[width=0.5\textwidth]{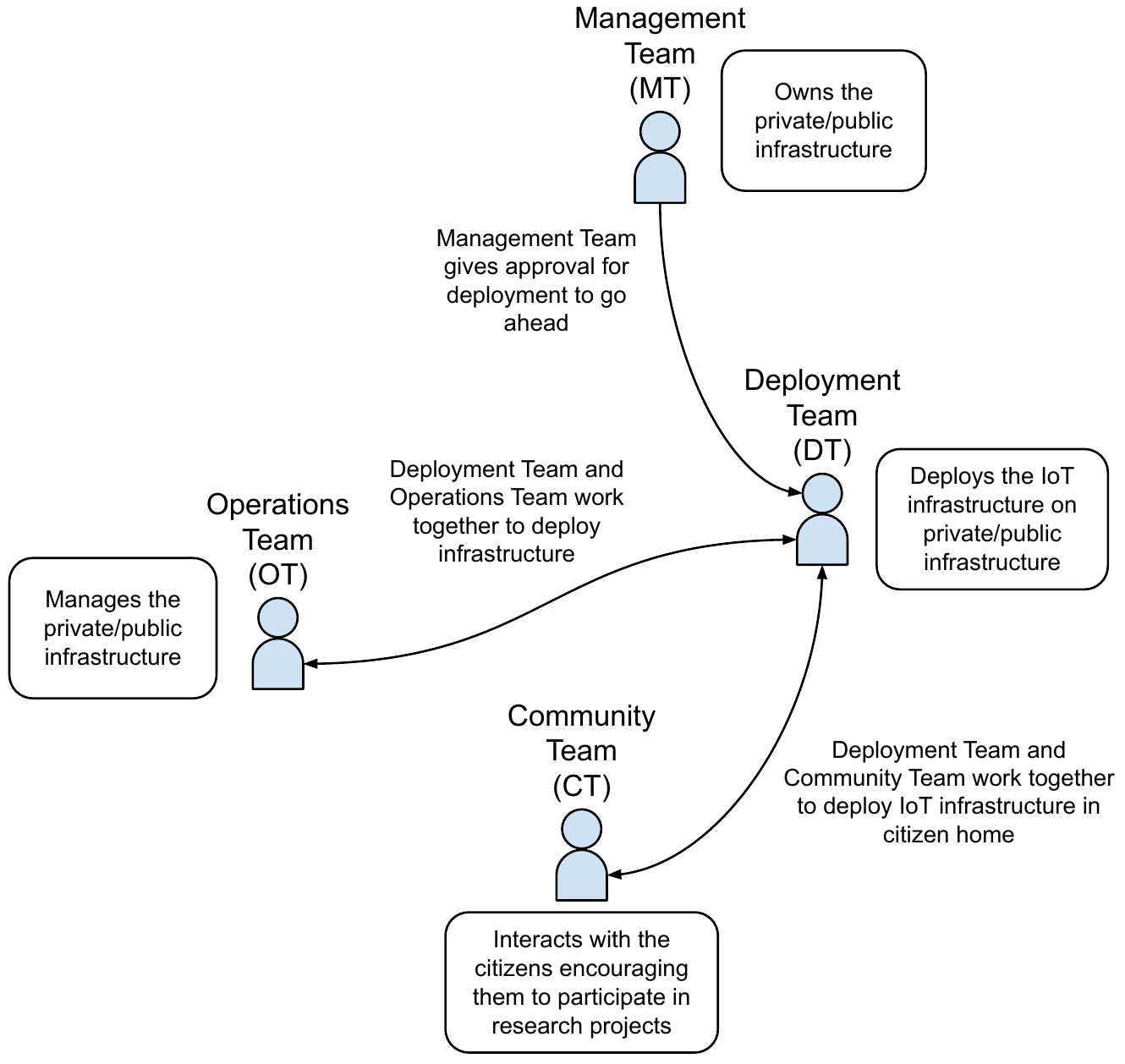}
    \caption{Different teams involved in smart city research projects and their relationships}
    \label{fig:team_relationships}
\end{figure}

Research projects often require the installation of sensors in the environment/infrastructure owned by a different party. However, before deploying a large-scale deployment, it is important to have a small-scale deployment to understand real-world challenges and build confidence with infrastructure owners. Devices may behave differently depending on external factors (power supply, network infrastructure, and physical environment)~\cite{Kurkovsky2017, Jackson2017}. The small-scale deployment could include one citizen house, streetlamp or vehicle. Deploying scientific infrastructure on others infrastructure (bridge - owned by a trust, streetlamps - owned by the council, citizens' house - rented or owned by tenants) requires partnership with the respective owner~\cite{Catlett2017}. There could be two individual bodies governing the infrastructure, first, the management team (MT) (board of directors, members of C-suite) and second, the operations teams (OT) (people managing/implementing the infrastructure). We refer to the research team (the team that deploys the infrastructure) as DT for brevity. \autoref{fig:team_relationships} provides the different teams and their relationships.

During multiple projects involving device deployments, it was found that it is essential to gain the MT's trust (such as citizens and the city council) and inform them about the benefits of deploying the monitoring infrastructure. They will require assurance that the DT takes their work seriously and that installing the monitoring infrastructure will not disrupt their infrastructure working in any way.

Once the MT is on board, the DT must work with the OT. OT could be performing essential jobs such as keeping the city, a bridge running or operating their electric bicycle platform. The OT of different companies has their own key performance indicators (KPIs), processes, and structures. The challenge for DT personnel is to fit into that culture without causing problems. The DT should provide details (make, models, working, safety, security) of the monitoring infrastructure to gain OT's confidence and trust. The DT should experiment with the OT infrastructure without disrupting them and not being a burden. They need to explain and provide realistic expectations about the research project and what and how they will be doing it. Furthermore, the relationship between DT and OT should be sufficiently positive so that the research team can fit the practise of the infrastructure operations team and that OT is happy to work with DT.

Finally, the DT should behave safely, securely, and carefully while working with the OT. The DT must be aware of health and safety concerns~\cite{Jackson2017} and respect other people's time. For example, installing sensors on other infrastructures is often cancelled for non-technical reasons (e.g. violating health and safety requirements). Installing the sensors on an initial site (first house, streetlamp) will build up the DT's confidence and relationship with the OT/MT team.


\subsection{Implementation/Deployment (in the real world)}

Data-gathering research infrastructure can be deployed at citizens' houses, private buildings (offices), and public places (streetlamps, council vehicles). All have a different set of challenges. First, we cover the challenges faced in the deployment in citizen homes and public spaces. In addition to the deployment team (DT), we denote the community team interacting with citizens as CT. CT is often responsible for interacting with citizens and informing them about the project research objectives and results. They are the bridge between citizens and the DT.

From the perspective of citizen participation, privacy and transparency, it is also a good idea to display the data the device collects and how it is used by providing documentation near the device~\cite{emami2021privacy, Jackson2017}. It is also important to mention to whom the device belongs and where to contact for more information~\cite{Barrenetxea2008}.

\subsubsection*{\textbf{Deployment in citizen houses}}
Challenges faced by the CT can be divided into \textbf{i.} finding a way to interact with citizens \textbf{ii.} encouraging and involving them to participate in the research project \textbf{iii.} providing adequate information to citizens \textbf{iv.} maintaining regular contact with citizens.
\newline
\newline
\noindent\textbf{Finding potential motivated citizens:} Recruitment and engagement of citizens (potentially motivated) is challenging, requires proper planning and often requires plenty of time. It is more manageable in areas with community cohesion or a coordinating body to promote the project~\cite{Porto2013}. Recruitment works best using various methods, from brochures and social media to door-knocking and face-to-face visits~\cite{3ebest}. While interacting with citizens during the REPLICATE, Twinergy project, it was found that it is essential to consider literacy rates within the pilot area and to publish information/leaflets in the local language~\cite{3ebest} for non-native English citizens. Also, over the years, the CT often knows citizens from previous engagements who would be happy to participate. Local events are a good way to attract interest. The CT organises small events or has a booth with information during open markets. Before engagement, it is essential to check whether there is a specific research project requirement, such as the deployment of devices in citizens' houses with diabetes or Parkinson's disease or citizens with solar PV or in an excellent socio-economical situation~\cite{twinergy_website}. In such cases, CT interacts with different community groups through local community centres and social media applications, such as Facebook and Nextdoor~\cite{nextdoor_website}. Additionally, pandemic events such as COVID-19 make it difficult for CT to interact with citizens.

After identifying the recruitment method to build citizen interest, it is essential to consider the larger picture and connect people to these concepts. The CT also uses creativity and art to get that message across. The involvement of the physical and kinaesthetic aspects of the citizen often helps people become more involved, engaged, and excited about the research project. For example, \ac{KWMC} CT installed a booth with a workshop of crafts activities to engage citizens during an open market. Once citizens are engaged and enjoying the craft activities, the CT asks for details about where they live and introduces the research project objectives. Additionally, citizens often drop out of the research study for multiple reasons, such as ill health, changes in circumstances, moving house, and occasional frustration with technology/process~\cite{Porto2013}. Therefore, having more participants than the project requires and having few citizens as a reserve is always good.
\newline
\newline
\textbf{Citizen encouragement:}
The second challenge of CT is to get citizens excited about the project. It often comes to a fundamentally simple proposition: why they (citizens) would get involved and what is in it for them. Citizen participation becomes more complicated if the project requires a power supply or Wi-Fi (which costs money to citizens). When expenses are covered, there will still be a disruption in citizen life due to the installation of devices in houses~\cite{Lundrigan2019}. In many cases, incentives (free Wi-Fi access, free tablets, shopping vouchers, or the opportunity to win a smartphone) will not convince citizens to participate. It is essential to think carefully about how citizens can be recruited and maintain interest among them~\cite{3ebest}. For many people, simply getting involved is a barrier. For example, Twinergy~\cite{twinergy_website} requires that citizens have solar PV connected to their homes. However, citizens who have solar PV will be early adopters and tech-savvy, so they may not be interested in the project. Citizen onboarding to the research project is challenging and can involve different efforts depending on citizens' eagerness and benefits.
\newline
\newline
\textbf{Respecting citizen time and preferences:}
Deploying the endpoints in a home involves connecting up the sensors (using Wi-Fi, LPWAN or mesh networks). It can take a reasonable time, depending on the number of endpoints configured or connected and finding and deciding on a suitable place to keep the device, talk to the participants, and answer their questions~\cite{Lundrigan2019}. Technology that is easy to install with little or no cabling is preferred. Radio transmission devices are preferred as citizens do not prefer additional cables in their staircases and dwellings~\cite{3ebest}. During the Twinergy project, one participant decided not to install the technology because it would spoil their minimalist decor.

In case of Wi-Fi connectivity, DT would need the credentials (SSID and password) and can collect them through phone calls, online forms, or in-person. However, remotely managing the Wi-Fi credentials often results in issues such as participants being uncomfortable entering their password into a document, participants needing to know their Wi-Fi credentials, and mistakes made during communication (such as mistaking O with 0 (zero)). An incorrect Wi-Fi credential is only detected when the deployment occurs. In this case, the endpoints must be returned to the DT and loaded with the correct network name and password, or a visit to the participant's house is required to correct the credentials~\cite{Lundrigan2019}.

Furthermore, the endpoint devices must remain placed throughout the deployment period without damaging the participant's house (delicate surfaces such as precious antique wood and wallpaper)~\cite{beaudin2004lessons, Hnat2011}. It is advised to anticipate objects and environmental conditions that can affect installation. This includes moisture, the quality of surface finishes, the typical movement of the object, and the methods of interaction of inhabitants with the object~\cite{beaudin2004lessons, Sotres2017}. Often, the citizen, pet, or robot vacuum cleaner accidentally or unknowingly disconnects the power supply to the devices, causing a failure, resulting in loss of connectivity and data~\cite{Hnat2011}. Therefore, it is essential to identify the location of the device deployment at home carefully. The
DT must respect the citizen's house and time~\cite{Langendoen2006}. The longer the DT takes at a citizen's home, the more inconvenient it is for the citizen and their regular routine~\cite{Lundrigan2019}. Home visits of citizens for deployment and maintenance purposes must be highly optimised and efficient with preparation done beforehand~\cite{Hnat2011}.

Expecting user participation at all times is futile; expecting users to accurately record their activities for labelling data (such as who cooked dinner at what time) is challenging, as it requires citizens to remember and observe their lives~\cite{Hnat2011}.
\newline
\newline
\textbf{Device looks and deployment surrounding:}
User comfort, acceptance, and aesthetics of deployed devices are paramount for a successful deployment (especially for wearable endpoints or visible devices)~\cite{Fafoutis2017, Elsts2018, Chatterjee2018}. The citizen usually prefers the devices to look aesthetically or hidden away. When there are deployments in the citizen home, there must be no light emitting from devices deployed in bedrooms, as they can disturb users' sleep or affect user behavior~\cite{Fafoutis2017,beaudin2004lessons}. Furthermore, LEDs also consume a good amount of energy~\cite{Barrenetxea2008}. It would be good to have the ability in the endpoints to turn on/off the LEDs so that they can be on during debugging and off during real deployments~\cite{Oyedele2021}. For example, SCK deployed on the Cotham Hill citizen's house emits red light in case of setup issues; a senior resident was concerned and asked if it is safe to operate and has no fire hazard. In addition, it is essential to ensure that the device does not make any noise that can affect the lives of citizens~\cite{Hnat2011}.

It is also essential to note the device deployment conditions or the surrounding location to understand the sensor readings~\cite{beaudin2004lessons}. For example, a temperature reading in an area with direct sunlight will vary from a temperature reading in the shade~\cite{Catlett2017}. To provide another example, anomalies in the SCK noise sensor readings in the Cotham-Hill deployment were observed because of the direct sunlight on the SCK kit kept near the window. Direct sunlight leads to device heating and can affect sensor readings~\cite{Folea2020}. In public deployments, context is also essential (near an intersection, highway, garbage can, and recycling centres). It is critical to understand how local environmental conditions (indoor/outdoor/sunshine/rain/snow) will affect the deployments~\cite{Barrenetxea2008}.
\newline
\newline
\textbf{Citizens switching home broadband provider:}
The device installed in the house often connects to the Internet through the ethernet port of the broadband router or Wi-Fi (which requires broadband Wi-Fi credentials)~\cite{Yang2015, LeBlond2012}. For example, in REPLICATE, the endpoint connects to the edge device using ethernet to forward and route all traffic from \ac{VPN} to the smart city platform. Most edge devices are SBCs with one ethernet port and a Wi-Fi adapter. Therefore, when the Ethernet port is occupied, the device must connect via Wi-Fi to connect to the Internet.

Citizens often change their broadband providers from one to six months to a year, leading to the change of Wi-Fi credentials (SSID and password) and loss of Internet connectivity and data. The DT does not have any mechanism to replace the Wi-Fi passphrase but requests the household owners to change the Wi-Fi passwords to what it was before, including the SSID, so that the device can connect to the Wi-Fi network. The other way is to plug the edge device into a monitor, attach a keyboard/mouse, provide credentials to the household owner and ask them to run the script to change the Wi-Fi password. However, most homeowners are not tech-savvy, making changes difficult. In addition, many citizens are unfamiliar with the technology introduced to their homes. For example, citizens might not have the experience of using a tablet or have problems accessing their information via the Internet~\cite{Porto2013}.

\subsubsection*{\textbf{Deployment in private building and public spaces}}

The deployment of any devices on the city's infrastructure (buses, garbage trucks, streetlamps) requires the willingness and collaboration of the city council~\cite{Sotres2017}. Similarly, deploying devices on private buildings requires the building management team's approval. During the Clifton Suspension Bridge project, it was found that it is essential to ensure that any device deployed does not hinder the functioning of city infrastructure or private buildings. The power source for the deployed device must be planned (such as streetlamp power or car batteries when deployed on buses/trucks, mains powered, battery powered)~\cite{Sotres2017}. It takes time and effort to secure permissions with the relevant infrastructure owners to deploy devices. Therefore, it is essential to identify the locations with the most significant impact to deploy the edge/endpoint that provides the most value to the stakeholders of the research/project~\cite{Lundrigan2019, Dehwah2015}. Suppose the device is deployed on the streetlamps and contains a downward camera. In that case, it might be a good idea to mount the device at a higher position to protect it from vandalism or theft~\cite{Dehwah2015, Basnyat2020}. This would also allow an extensive view from the camera, allowing images of the entire intersection/park.

For a successful public deployment of infrastructure, policies, agreements, processes, public engagement, and interactions are necessary.

\textbf{Public engagement:} Public engagement is essential for the success of the research project. It brings city residents closer to the project and makes them active participants. It helps citizens without technology experience to discuss and learn the use of data and technology. This broader citizenry can explore and develop solutions to urban issues by proposing ideas for how collected data can be used. Community centres or community outreach help to publicise the project. There must be a named person to whom participants can go with any questions~\cite{3ebest}. Face-to-face meetings help people identify and assign a named person to a project. Throughout the project, excellent and responsive personal support from a friendly and accessible coordinator (in the form of a building manager, a housing association contact, or even a community leader) can increase engagement. Any research project aiming to impact citizens' lives or affect behaviour change must build a relationship with participants and a deep understanding of their contexts and motivations to increase engagement and participation levels. Users must feel involved in each stage of project development and see that their participation is valued and that their input can have a real impact~\cite{3ebest}. In addition, periodic reinforcement of the message and encouragement by contact between the neighbours and the central coordinator helps keep the motivation and the participants interested~\cite{Porto2013}. It is vital to provide ongoing support through visits, calls, and workshops, especially for those who find technology difficult or have literacy problems. Creating a relationship with participants based on trust and responsibility for communicating bad and good news~\cite{3ebest} helps the researcher and the citizen.

Also, there is a possibility that the research projects engage with people from underserved or disadvantaged socio-economic or minority ethnic backgrounds. It is crucial not to lump them into one group. The CT must treat everyone equally and ensure that communication with the citizens is appropriate and accessible, and no one should be offended.

Furthermore, the amount of information must be provided in an easily digestible fashion (short video, infographics, a mechanism with which citizens can engage and interact) to get comfortable with the idea and not overwhelm them. The research project results depend heavily on the interaction and feedback of the participants. Hence, it is essential to ensure that easy-to-understand and straightforward messages are used to communicate with citizens (communication is key)~\cite{Porto2013}. For example, SPHERE created a 3 min animated video~\cite{sphere_youtube_website} to provide information to the citizens. Being active on social media, such as Twitter, responding to media requests for interviews, and publishing detailed information about the research project on the website/pamphlets/leaflets helps improve public perception and participation~\cite{Committee2018}.

In the case of deployment in citizen houses, once citizens are on board and have signed the consent forms (ensures commitment and guarantees confidentiality), and the DT has installed the devices in their house, it is still essential to maintain regular contact with the citizens to ensure devices are working and they can use the technology and data provided for their benefit. Another minor challenge for CT is managing the signed consent forms provided to citizens for participation. Encouraging all participants to return completed questionnaires is always challenging and must be considered for any citizen attitude/behavioural analysis~\cite{Porto2013}.

\textbf{Transparency:}
Deploying any public infrastructure requires transparency, privacy protection, and system security. The public usually suspects publicly deployed devices based on fears about surveillance and data collected by the node~\cite{Jackson2017}. It is essential to develop and provide privacy and governance policies to show the project's commitment to transparency and privacy. The privacy policy should provide what data are being collected, processed, used, destroyed, or made available to city residents. Additionally, allowing open comments from the citizens and community on the policy drafts help gain citizen confidence. DT/CT can arrange community meetings for citizens to ask questions about the draft policies. It is essential to resolve all the comments and questions publicly, consider citizen feedback for policy revision, and include a report of the public engagement process. The public/government cybersecurity centre can assess the deployed system security and privacy practices to ensure system security and gain public trust~\cite{Committee2018}.

In the case of deployment at home, citizens will have questions about the different endpoints, frequencies used, data collected, and how data will be used~\cite{Yang2015} and stored. On the contrary, the DT requests information from the CT on the house floor plan to design/customise the sensors according to the requirements~\cite{Lundrigan2019, Hnat2011}. The above situation can often land the CT in a dilemma, as projects often decide which sensors will be deployed and data collected late in the project. Furthermore, the CT cannot tell the citizens about the sensors until the project's data and requirements are well defined. Citizens can only decide whether they want to participate in the project once they have clarity on what is collected, which means that the CT cannot provide house details to the DT. Therefore, it is better to perform a requirement analysis (\autoref{subsec:challenges:requirement}) earlier in the project to understand data collection and be transparent with citizens.

\subsection{Operational Challenges}

Research projects also have operational challenges, which are problems that arise and can render a project less efficient.

\subsubsection*{\textbf{Skills shortage}}
A significant challenge is the shortage of people with the appropriate skill set to act as system architects in urban monitoring research projects. Research projects (a collaboration between universities, industry, and city councils) are often for 1-5 years. The people who develop and manage the urban monitoring platform are research associates and doctoral students, who mainly cover only part of the required skill set. Furthermore, students who maintain the project often work part-time due to semesters and other courses, leading to staffing problems~\cite{Gvk2019}. Experience and knowledge in system administration, cloud infrastructure, networking, DevOps, and cybersecurity are required~\cite{Keahey2019, Sony2020, ballon2011there}.

\subsubsection*{\textbf{Different expectations and goals}}
Research projects can have multiple partners and collaborations. Each partner can have a different set of expertise, business models, expectations and their own project agenda on how it benefits them~\cite{Sony2020}. There may be cases where collaboration priorities are different, which can create challenges in communication and work completion. Teamwork is essential for project success~\cite{Kurkovsky2017, Sony2020}.
Furthermore, research members can have other KPIs on which their managers judge their performance. If the delivery of the research project is not one of them, it can affect the researcher's commitment to the project. There will always be members in the project who will be hard working, average working, and who would cause trouble; always good to identify the right person for the right work.

\subsubsection*{\textbf{Clear, concise communication}}

Research projects often include multiple meetings to discuss various objectives and goals of the project. It is crucial to have clearly defined agendas and final takeaways. Also, it is a good practice to invite only a few key people or technical leads to the meeting for clear and concise communication. In addition, face-to-face meetings are preferred over online discussions, especially brainstorming sessions. Things become delayed if the parties involved do not communicate clearly and concisely.

\subsubsection*{\textbf{Risk Management}}
The research project should also have risk management that considers different issues in the project schedule. Risks could include COVID-19 affecting people, datasets not available for analysis, delays in setting up the testbeds, deployment of devices in public spaces, and related safety issues (electrical hazards, devices falling from streetlamps), among others. Furthermore, it should include critical personal backup plans if someone gets sick or leaves the project/company.
Furthermore, suppose that the deployed devices are expected to work after the end of the research phase. In that case, it is essential to have a handover-takeover (HOTO) (including hiring and transferring skills) to continue a successful project. Often, the platform and devices require some human intervention to operate~\cite{Jackson2017}.

\subsubsection*{\textbf{Infrastructure availability}}
There will be inevitable situations outside the control of the research team. For example, infrastructure suffers from an outage, a global internet outage, or installed devices affected by weather~\cite{Jackson2017}. As another example, there is little the DT can do if the cloud tier is hosted on city-council infrastructure and an outage occurs with their main administrator on leave. Case in point, the Internet recently suffered a significant outage of approximately one hour~\cite{Internetoutage}, leaving multiple cloud services unavailable.

Devices required for deployment must be purchased early. Importing devices from another country and connecting them to the home network is expensive and challenging. A significant amount of time is lost in the shipment of devices across continents, exacerbated by having to work in multiple timezones~\cite{Basnyat2020}.

\subsubsection*{\textbf{Partnerships}}
It is essential to have the support and partnership of the city council~\cite{Lago2021}. The city council officials can act as a catalyst for informing and organising discussions with other city departments (electricity board, hospitals, recycling). These other departments can update the city council about the project and ask for their input on deployment locations or how the project can support a particular department in solving its challenges. The research project, depending on its objective, can support the vision of the city plan (usually published year-by-year, such as the Bristol city plan~\cite{BristolCityCouncil2020}, Belfast Agenda~\cite{BelfastCItyCouncil2014}, Chicago Technology plan~\cite{TheCityofChicago2013}) in terms of how the research project and the deployment of the public infrastructure can allow the city to use technology and data for engagement, innovation, inclusion, and opportunity.

In addition, it is essential to engage and win the confidence of city departments and employees by involving them in the project. For example, suppose that the infrastructure will be installed on city streetlamps. In that case, it is important to bring prototype units to the electrical department and seek their input on electrical safety and mounting procedure, effectively gaining their confidence and working as a team toward a common goal.

\subsubsection*{\textbf{Logistics}}

The DT should be aware of the design of the nodes, the installation procedures, the node deployment locations, and other information. In addition, they should have ownership and power to make decisions on the fly, such as moving a node to a different street corner due to a blocked view during installation. Interactions and conversations can lead to collaborations and understanding of how research data collected by public deployment can be used and integrated into existing city data platforms (such as Bristol Open Data~\cite{bristol_open_data_website}, London Datastore~\cite{london_open_data_website}).

Furthermore, DT can create communication channels such as surveys and forms to collect the location of the node deployment, the type of data, and the problems to be solved from the project stakeholders, city departments, communities, research groups, and residents~\cite{Committee2018}.

%% file: Diagrams/exposed-services-thread-model.latex
\definecolor{plantucolor0000}{RGB}{0,0,0}
\definecolor{plantucolor0001}{RGB}{229,230,229}
\definecolor{plantucolor0002}{RGB}{226,226,240}
\definecolor{plantucolor0003}{RGB}{24,24,24}
\definecolor{plantucolor0004}{RGB}{0,47,95}
\definecolor{plantucolor0005}{RGB}{241,241,241}
\begin{tikzpicture}[yscale=-1
,pstyle0/.style={color=plantucolor0001,fill=plantucolor0001,line width=1.0pt}
,pstyle1/.style={color=plantucolor0003,fill=plantucolor0002,line width=1.0pt}
,pstyle3/.style={color=plantucolor0003,line width=1.0pt}
,pstyle4/.style={color=plantucolor0003,fill=plantucolor0005,line width=0.5pt}
]
\node at (5pt,5pt)[below right,color=black]{Internet};
\node at (29.4155pt,136.9063pt)[below right,color=black]{Edge};
\node at (12.5353pt,281.6172pt)[below right,color=black]{LPWAN};
\draw[pstyle0] (118.0351pt,177.793pt) rectangle (329.3962pt,235.7305pt);
\node at (123.0351pt,177.793pt)[below right,color=black]{Edge};
\draw[pstyle0] (236.1235pt,39.4844pt) rectangle (330.6689pt,97.4219pt);
\node at (241.1235pt,39.4844pt)[below right,color=black]{Cloud};
\draw[pstyle1] (69.68pt,9.4844pt) rectangle (340.6689pt,14.4844pt);
\draw[color=plantucolor0003,fill=plantucolor0004,line width=1.0pt] (69.68pt,141.3906pt) rectangle (340.6689pt,146.3906pt);
\draw[pstyle1] (69.68pt,286.1016pt) rectangle (340.6689pt,291.1016pt);
\draw[pstyle3] (149.9017pt,14.4844pt) -- (149.9017pt,53.4531pt);
\draw[pstyle3] (149.9017pt,97.4219pt) -- (149.9017pt,141.3906pt);
\draw[pstyle3] (285.3962pt,14.4844pt) -- (285.3962pt,58.4531pt);
\draw[pstyle3] (145.9017pt,146.3906pt) -- (145.9017pt,196.7617pt);
\draw[pstyle3] (281.3962pt,146.3906pt) -- (281.3962pt,196.7617pt);
\draw[pstyle3] (281.3962pt,230.7305pt) -- (281.3962pt,286.1016pt);
\node at (247.5503pt,252.0137pt)[below right,color=black]{SLIP Radio};
\draw[pstyle3] (149.9017pt,291.1016pt) -- (149.9017pt,327.5039pt);
\node at (104.4119pt,300.4004pt)[below right,color=black]{Endpoint Radio};
\draw[pstyle4] (109.0243pt,63.0941pt) ..controls (111.3826pt,56.4472pt) and (115.612pt,55.2112pt) .. (121.1649pt,59.5703pt) ..controls (123.9335pt,51.197pt) and (129.9859pt,51.1874pt) .. (135.4758pt,56.7824pt) ..controls (138.8142pt,49.3671pt) and (144.6558pt,48.9562pt) .. (149.5123pt,55.2252pt) ..controls (155.4497pt,50.001pt) and (159.9556pt,51.1959pt) .. (163.7196pt,57.8292pt) ..controls (169.5211pt,53.3681pt) and (173.5178pt,54.5739pt) .. (176.5021pt,61.1051pt) ..controls (181.3596pt,55.0077pt) and (186.1465pt,54.9037pt) .. (189.8475pt,62.2529pt) ..controls (198.483pt,62.5208pt) and (203.0791pt,66.887pt) .. (199.3476pt,75.8168pt) ..controls (203.3926pt,86.4071pt) and (198.8774pt,90.0815pt) .. (188.847pt,90.8218pt) ..controls (184.6615pt,97.6204pt) and (179.9123pt,98.2468pt) .. (175.369pt,91.1692pt) ..controls (171.0456pt,97.9342pt) and (166.879pt,96.6274pt) .. (162.4953pt,91.2331pt) ..controls (158.5445pt,97.8212pt) and (153.6801pt,96.7298pt) .. (149.4959pt,91.4897pt) ..controls (144.283pt,98.3785pt) and (139.3035pt,98.256pt) .. (134.5948pt,90.96pt) ..controls (131.3428pt,96.5508pt) and (124.6181pt,98.8248pt) .. (120.919pt,91.4309pt) ..controls (116.1876pt,96.0467pt) and (111.7139pt,96.068pt) .. (108.4382pt,89.7667pt) ..controls (100.3396pt,88.2193pt) and (97.4291pt,84.5412pt) .. (100.7141pt,76.4104pt) ..controls (96.2272pt,68.8071pt) and (99.7927pt,62.1363pt) .. (109.0243pt,63.0941pt);
\node at (113.9933pt,68.4531pt)[below right,color=black]{LTE WAN};
\draw[pstyle4] (241.1235pt,58.4531pt) rectangle (325.6689pt,92.4219pt);
\node at (251.1235pt,68.4531pt)[below right,color=black]{endpoint};
\draw[pstyle4] (123.0351pt,196.7617pt) rectangle (172.7684pt,230.7305pt);
\node at (133.0351pt,206.7617pt)[below right,color=black]{LAN};
\draw[pstyle4] (242.3962pt,196.7617pt) rectangle (324.3962pt,230.7305pt);
\node at (252.3962pt,206.7617pt)[below right,color=black]{RPL/IPv6};
\draw[pstyle4] (84.68pt,327.5039pt) rectangle (211.1235pt,361.4727pt);
\node at (94.68pt,337.5039pt)[below right,color=black]{Endpoint Node};
\end{tikzpicture}

%% file: Diagrams/3tier-wsn-data-path.latex
\definecolor{plantucolor0000}{RGB}{173,216,230}
\definecolor{plantucolor0001}{RGB}{24,24,24}
\definecolor{plantucolor0002}{RGB}{0,0,0}
\definecolor{plantucolor0003}{RGB}{144,238,144}
\definecolor{plantucolor0004}{RGB}{255,255,255}
\definecolor{plantucolor0005}{RGB}{226,226,240}
\definecolor{plantucolor0006}{RGB}{238,238,238}
\begin{tikzpicture}[yscale=-1
,pstyle0/.style={color=plantucolor0001,fill=plantucolor0000,line width=0.5pt}
,pstyle2/.style={color=black,fill=plantucolor0004,line width=1.5pt}
,pstyle3/.style={color=plantucolor0001,line width=0.5pt,dash pattern=on 5.0pt off 5.0pt}
,pstyle4/.style={color=plantucolor0001,fill=plantucolor0005,line width=0.5pt}
,pstyle5/.style={color=black,fill=plantucolor0006,line width=1.5pt}
,pstyle6/.style={color=black,line width=1.5pt}
,pstyle7/.style={color=plantucolor0001,fill=plantucolor0001,line width=1.0pt}
,pstyle8/.style={color=plantucolor0001,line width=1.0pt}
]
\draw[pstyle0] (16pt,6pt) rectangle (198.6273pt,292.082pt);
\node at (68.1379pt,6pt)[below right,color=black]{\textbf{Endpoint}};
\draw[pstyle0] (235.5116pt,6pt) rectangle (360.9738pt,292.082pt);
\node at (276.8649pt,6pt)[below right,color=black]{\textbf{Edge}};
\draw[color=plantucolor0001,fill=plantucolor0003,line width=0.5pt] (421.6159pt,6pt) rectangle (516.3534pt,292.082pt);
\node at (444.6286pt,6pt)[below right,color=black]{\textbf{Cloud}};
\draw[pstyle2] (10pt,73.5693pt) rectangle (204.6273pt,119.9746pt);
\draw[pstyle2] (127.849pt,133.9746pt) rectangle (300.1955pt,180.3799pt);
\draw[pstyle2] (290.1955pt,194.3799pt) rectangle (522.3534pt,240.7852pt);
\draw[pstyle3] (68pt,56.5693pt) -- (68pt,257.7852pt);
\draw[pstyle3] (165.849pt,56.5693pt) -- (165.849pt,257.7852pt);
\draw[pstyle3] (264.5116pt,56.5693pt) -- (264.5116pt,257.7852pt);
\draw[pstyle3] (328.1955pt,56.5693pt) -- (328.1955pt,257.7852pt);
\draw[pstyle3] (468.6159pt,56.5693pt) -- (468.6159pt,257.7852pt);
\draw[pstyle4] (20pt,30.2725pt) arc (180:270:5pt) -- (25pt,25.2725pt) -- (111.2301pt,25.2725pt) arc (270:360:5pt) -- (116.2301pt,30.2725pt) -- (116.2301pt,50.5693pt) arc (0:90:5pt) -- (111.2301pt,55.5693pt) -- (25pt,55.5693pt) arc (90:180:5pt) -- (20pt,50.5693pt) -- cycle;
\node at (27pt,32.2725pt)[below right,color=black]{Transducer};
\draw[pstyle4] (20pt,261.7852pt) arc (180:270:5pt) -- (25pt,256.7852pt) -- (111.2301pt,256.7852pt) arc (270:360:5pt) -- (116.2301pt,261.7852pt) -- (116.2301pt,282.082pt) arc (0:90:5pt) -- (111.2301pt,287.082pt) -- (25pt,287.082pt) arc (90:180:5pt) -- (20pt,282.082pt) -- cycle;
\node at (27pt,263.7852pt)[below right,color=black]{Transducer};
\draw[pstyle4] (137.849pt,30.2725pt) arc (180:270:5pt) -- (142.849pt,25.2725pt) -- (189.6273pt,25.2725pt) arc (270:360:5pt) -- (194.6273pt,30.2725pt) -- (194.6273pt,50.5693pt) arc (0:90:5pt) -- (189.6273pt,55.5693pt) -- (142.849pt,55.5693pt) arc (90:180:5pt) -- (137.849pt,50.5693pt) -- cycle;
\node at (144.849pt,32.2725pt)[below right,color=black]{MCU};
\draw[pstyle4] (137.849pt,261.7852pt) arc (180:270:5pt) -- (142.849pt,256.7852pt) -- (189.6273pt,256.7852pt) arc (270:360:5pt) -- (194.6273pt,261.7852pt) -- (194.6273pt,282.082pt) arc (0:90:5pt) -- (189.6273pt,287.082pt) -- (142.849pt,287.082pt) arc (90:180:5pt) -- (137.849pt,282.082pt) -- cycle;
\node at (144.849pt,263.7852pt)[below right,color=black]{MCU};
\draw[pstyle4] (239.5116pt,30.2725pt) arc (180:270:5pt) -- (244.5116pt,25.2725pt) -- (285.1955pt,25.2725pt) arc (270:360:5pt) -- (290.1955pt,30.2725pt) -- (290.1955pt,50.5693pt) arc (0:90:5pt) -- (285.1955pt,55.5693pt) -- (244.5116pt,55.5693pt) arc (90:180:5pt) -- (239.5116pt,50.5693pt) -- cycle;
\node at (246.5116pt,32.2725pt)[below right,color=black]{LAN};
\draw[pstyle4] (239.5116pt,261.7852pt) arc (180:270:5pt) -- (244.5116pt,256.7852pt) -- (285.1955pt,256.7852pt) arc (270:360:5pt) -- (290.1955pt,261.7852pt) -- (290.1955pt,282.082pt) arc (0:90:5pt) -- (285.1955pt,287.082pt) -- (244.5116pt,287.082pt) arc (90:180:5pt) -- (239.5116pt,282.082pt) -- cycle;
\node at (246.5116pt,263.7852pt)[below right,color=black]{LAN};
\draw[pstyle4] (300.1955pt,30.2725pt) arc (180:270:5pt) -- (305.1955pt,25.2725pt) -- (351.9738pt,25.2725pt) arc (270:360:5pt) -- (356.9738pt,30.2725pt) -- (356.9738pt,50.5693pt) arc (0:90:5pt) -- (351.9738pt,55.5693pt) -- (305.1955pt,55.5693pt) arc (90:180:5pt) -- (300.1955pt,50.5693pt) -- cycle;
\node at (307.1955pt,32.2725pt)[below right,color=black]{WAN};
\draw[pstyle4] (300.1955pt,261.7852pt) arc (180:270:5pt) -- (305.1955pt,256.7852pt) -- (351.9738pt,256.7852pt) arc (270:360:5pt) -- (356.9738pt,261.7852pt) -- (356.9738pt,282.082pt) arc (0:90:5pt) -- (351.9738pt,287.082pt) -- (305.1955pt,287.082pt) arc (90:180:5pt) -- (300.1955pt,282.082pt) -- cycle;
\node at (307.1955pt,263.7852pt)[below right,color=black]{WAN};
\draw[pstyle4] (425.6159pt,30.2725pt) arc (180:270:5pt) -- (430.6159pt,25.2725pt) -- (507.3534pt,25.2725pt) arc (270:360:5pt) -- (512.3534pt,30.2725pt) -- (512.3534pt,50.5693pt) arc (0:90:5pt) -- (507.3534pt,55.5693pt) -- (430.6159pt,55.5693pt) arc (90:180:5pt) -- (425.6159pt,50.5693pt) -- cycle;
\node at (432.6159pt,32.2725pt)[below right,color=black]{Endpoint};
\draw[pstyle4] (425.6159pt,261.7852pt) arc (180:270:5pt) -- (430.6159pt,256.7852pt) -- (507.3534pt,256.7852pt) arc (270:360:5pt) -- (512.3534pt,261.7852pt) -- (512.3534pt,282.082pt) arc (0:90:5pt) -- (507.3534pt,287.082pt) -- (430.6159pt,287.082pt) arc (90:180:5pt) -- (425.6159pt,282.082pt) -- cycle;
\node at (432.6159pt,263.7852pt)[below right,color=black]{Endpoint};
\draw[pstyle5] (10pt,73.5693pt) -- (157.7448pt,73.5693pt) -- (157.7448pt,80.8418pt) -- (147.7448pt,90.8418pt) -- (10pt,90.8418pt) -- (10pt,73.5693pt);
\draw[pstyle6] (10pt,73.5693pt) rectangle (204.6273pt,119.9746pt);
\node at (25pt,74.5693pt)[below right,color=black]{\textbf{Digitization}};
\draw[pstyle7] (154.2381pt,107.9746pt) -- (164.2381pt,111.9746pt) -- (154.2381pt,115.9746pt) -- (158.2381pt,111.9746pt) -- cycle;
\draw[pstyle8] (68.1151pt,111.9746pt) -- (160.2381pt,111.9746pt);
\node at (75.1151pt,94.8418pt)[below right,color=black]{Raw Data};
\draw[pstyle5] (127.849pt,133.9746pt) -- (235.2786pt,133.9746pt) -- (235.2786pt,141.2471pt) -- (225.2786pt,151.2471pt) -- (127.849pt,151.2471pt) -- (127.849pt,133.9746pt);
\draw[pstyle6] (127.849pt,133.9746pt) rectangle (300.1955pt,180.3799pt);
\node at (142.849pt,134.9746pt)[below right,color=black]{\textbf{LPWAN}};
\draw[pstyle7] (252.8535pt,168.3799pt) -- (262.8535pt,172.3799pt) -- (252.8535pt,176.3799pt) -- (256.8535pt,172.3799pt) -- cycle;
\draw[pstyle8] (166.2381pt,172.3799pt) -- (258.8535pt,172.3799pt);
\node at (173.2381pt,155.2471pt)[below right,color=black]{MAC/PHY};
\draw[pstyle5] (290.1955pt,194.3799pt) -- (407.0184pt,194.3799pt) -- (407.0184pt,201.6523pt) -- (397.0184pt,211.6523pt) -- (290.1955pt,211.6523pt) -- (290.1955pt,194.3799pt);
\draw[pstyle6] (290.1955pt,194.3799pt) rectangle (522.3534pt,240.7852pt);
\node at (305.1955pt,195.3799pt)[below right,color=black]{\textbf{Internet}};
\draw[pstyle7] (456.9847pt,228.7852pt) -- (466.9847pt,232.7852pt) -- (456.9847pt,236.7852pt) -- (460.9847pt,232.7852pt) -- cycle;
\draw[pstyle8] (328.5847pt,232.7852pt) -- (462.9847pt,232.7852pt);
\node at (335.5847pt,215.6523pt)[below right,color=black]{Processed Data};
\end{tikzpicture}

%% file: Sections/6_conclusion.tex
The continued growth of wireless technologies has resulted in significant research into urban monitoring via data-gathering IoT testbeds. These research testbeds follow a typical three-tier architecture, and many designs and implementation challenges remain, including data privacy controls, network security, and device updates. We extracted these challenges and associated lessons learned by considering several real-world IoT testbed projects. We analyzed the projects in the context of the V-model development life cycle phases. We presented the project challenges and lessons learned organized by requirements analysis, system design, implementation, testing and deployment phases. We believe this will assist other urban monitoring researchers in planning future testbeds. We hope this research will prove valuable and reduce these projects' design and implementation costs.